\documentstyle[twoside,multicol,aps,epsf]{revtex}
\begin{document}
\author{J. P. Neirotti \\
and\\
David L. Freeman \\ Department of Chemistry, University of Rhode
Island\\
51 Lower College Road, Kingston, RI 02881-0809\\
and\\
J. D. Doll\\
Department of Chemistry, Brown University\\
Providence, RI 02912}
\title{The Approach to Ergodicity in Monte Carlo Simulations}
\date{\today}
\maketitle

\makeatletter
\global\@specialpagefalse
\def\@evenfoot{\thepage\hfill{\small\sf The Approach to Ergodicity in Monte Carlo Simulations}} 
\def\@oddfoot{
\ifnum\c@page>1
  {\small\sf J. P. Neirotti, D. L. Freeman. and J. D. Doll}\hfill\thepage
\fi
\ifnum\c@page=1
  \hfill\thepage\hfill\llap{\protect{PREPRINT}}
\fi
} 
\makeatother
\vspace{.5cm}
\begin{center}
{\bf Abstract}
\end{center}

\begin{abstract}
The approach to the ergodic limit in Monte Carlo simulations is
studied using both analytic and numerical methods.
With the help of a stochastic model, a metric is defined that enables the
examination of a simulation in both the ergodic and non-ergodic regimes.  
In the non-ergodic regime, the model implies how the simulation is expected
to approach ergodic behavior analytically, and the analytically inferred
decay law of the metric
allows the monitoring of the onset of ergodic behavior. The
metric is related to previously defined measures developed for molecular
dynamics simulations, and the metric
enables the comparison of the relative efficiencies of different
Monte Carlo schemes.
Applications to Lennard-Jones 13-particle clusters are shown
to match the model for
Metropolis, J-walking and parallel tempering based approaches.  The
relative efficiencies of these three Monte Carlo approaches are
compared, and the decay law is shown to be useful
in determining needed high temperature parameters in parallel tempering
and J-walking studies of atomic clusters.

\end{abstract}

\begin{multicols}{2}

\section{Introduction}

A goal of Monte Carlo (MC) simulations in statistical 
mechanics \cite{valleau} is the calculation of ensemble mean values of 
thermodynamic quantities. Ensemble mean values are multidimensional 
integrals over configuration space

\begin{equation}
\label{eq0}
\langle U\rangle = \int\!\!d{\bf x}\,P({\bf x})\,U({\bf x})\,\,,
\end{equation}

\noindent where $P({\bf x})$ is the probability of finding a system 
in the state defined by ${\bf x}$, and the functional form 
of $P({\bf x})$ depends on 
the ensemble investigated. MC simulations usually generate a sampling 
of configuration space $\left\{{\bf x}_k\right\}_{k=1}^K$ by the use of 
a stochastic process with stationary probability $P({\bf x}_k)$. The 
quantity $U$ evaluated at ${\bf x}_k$ is the output of the simulation 
$U({\bf x}_k)=U_k$, and its arithmetic mean value $\overline{U}$, 
in principle, must approach the ensemble mean value.\cite{valleau} 
In this paper we refer to the set of configurations generated in a Monte
Carlo simulation as a time sequence, and
we study the behavior of these temporal sequences $\left\{U_k\right\}$ and 
their arithmetic mean, to understand better how MC simulations approach 
ergodic behavior.  It
is important to emphasize that there are two time variables to consider.
The time variable $k$ labels the separate configurations generated in a
Monte Carlo walk.  Variations of properties with $k$ provide information
about the short-time behavior of a MC simulation.  The time
variable $K$ labels the total length of the MC walk, and variations of
computed properties with $K$ provide information about the convergence
of the simulation on a long time scale.

Given an infinite time, the stochastic walker in a MC simulation visits 
every allowed point in configuration space.\cite{wood} Ergodic behavior is reached 
when the length of the walk is sufficiently long to sample configuration 
space appropriately.\cite{thiru1} In practice, this does not mean that 
the space has been densely covered but that every region with non-negligible 
probability has been reached. In such a case
we can say that the simulation is effectively 
ergodic or that it has reached the ergodic limit.

For a finite walk, in the event of broken ergodicity \cite{palmer}, 
phase space is effectively disconnected. The different disconnected 
regions (called components) are separated by barriers of zero
effective probability. If a stochastic walker starts its walk in one 
of these regions, it may not cross the barriers within the time of the 
simulation. If the simulation length is increased, some barriers may become 
accessible for the walker and phase space is better sampled. We can 
conclude that a time $\tau$ exists such that, for simulation lengths shorter 
than $\tau$, the walker becomes trapped in one of the phase space components. 
For simulation lengths much larger than $\tau$, phase space is effectively 
covered by the walker.

In this study we imagine a system having more than one time scale
$\tau_1\ll\tau_2\ll\dots\ll\tau_{\Lambda}$. In a Monte Carlo simulation
each scale comes
from stochastic processes with different correlation times.\cite{gardiner}
A precise definition of the correlation times for Monte Carlo
processes is given in Section III, but for the moment we can think of
these correlation times as identical to physical time scales of the system
under study.  To
understand these time scales more fully,
it is useful to focus on an example.
Prototypical of systems having such disparate time scales are
atomic and molecular clusters.  Typical cluster potential surfaces have
many local minima separated by significant energy barriers. 
\cite{freeman1,wales1,wales2}
The local
minima can be grouped into basins of similar energies, with each basin
separated from other basins again by energy barriers.
At short
Monte Carlo times a cluster system executes small amplitude oscillations
about one of its potential minima.  We can think of these vibrational
time scales as the shortest time scales that define a cluster system. 
As the simulation time is extended the system eventually hops between different
local minima within the same basin.  
The time scale for the first hops between local minima
can be considered the next shortest time scale for the simulation.  At
still longer Monte Carlo times, the system hops between different energy
basins defining yet another time scale for the simulation.  This
grouping of time scales continues until the longest time scale for a
given system is reached.  At Monte Carlo times that are long compared to
this longest time scale, the simulation is ergodic.

Consider a system with several time scales as mentioned above. 
If the length of the simulation is smaller than the smallest correlation 
time, the walker may become trapped in an effectively disconnected region 
and the sampling of phase space is incomplete. By increasing the time, 
the memory of the initial condition in the sampling decreases as 
the walker crosses to other previously unreachable regions. These oscillations
and hoppings can be modeled by a superposition of stochastic processes with 
different correlation
times. These processes with non-zero correlation times are known as {\em 
colored noise} processes (as opposed to zero correlation time {\em white 
noise} processes). \cite{gardiner} From the study of the autocorrelation 
functions of a stochastic model defined using these colored noise processes, 
we can verify that, at a fixed run length $K$, there exist two different 
groups of processes; those that contribute to the autocorrelation function with
terms that decay like
$1/k$ (called {\em diffusive} processes), and those that contribute to the 
autocorrelation function with terms that decay slower than $1/k$ (called {\em 
non-diffusive} processes). When the time of the simulation is increased, some
non-diffusive processes at shorter run lengths, start to contribute to the
autocorrelation function like diffusive processes. After the walk length reaches
the largest correlation time $\tau_{\Lambda}$, all processes contribute to the
autocorrelation function with terms that decay like $1/k$. At this point, the
simulation is at the diffusive regime and effective ergodicity has been reached.
A principal
goal of this work is 
to investigate the way in which the MC output $\left\{U_k\right\}$ reaches 
the diffusive limit (i.e. the ergodic limit) by studying the properties 
of autocorrelation functions under changes of scale in time, 
$K\to bK$ with $b>1$. 
By time scaling 
it is possible to infer the decay law of the non-diffusive contributions
with respect to the total simulation
time $K$.  The functional dependence of the non-diffusive contributions
on the parameter $b$ that is used to scale $K$ is determined empirically.
We have found the decay law so determined 
to be a particularly valuable method of concluding when a simulation
can be considered ergodic.  Unlike previous studies
\cite{thiru1,thiru2,thiru3,thiru4}
that only have
investigated the behavior of certain autocorrelation functions in the
ergodic regime, by focusing on the approach to ergodic behavior we have
a more careful monitor of the onset of ergodicity.
Once the non-diffusive 
contributions have decayed to a point where they are too small to be
distinguished from zero to within the fluctuations of the calculation,
we can say that the ergodic limit has been reached. 

The autocorrelation functions we use to measure the approach to the ergodic
limit are based on
one of the probes of ergodicity developed by Thirumalai and 
co-workers \cite{thiru1,thiru2,thiru3,thiru4}, 
and is often called the {\em energy metric}. 
The energy metric has been proposed as an alternative to other 
techniques \cite{thiru1} 
(like the study of the Lyapunov exponents \cite{lieberman}) for the study 
of ergodic properties in molecular dynamics (MD) simulations.  The
metric has been used to study the relative efficiency of MC simulation
methods as well. \cite{straub}
The MC metric as used in the current work 
can easily be extended from the energy to
other scalar observables of the system. 

We present two key issues in this paper. First, from the knowledge of the 
decay law of the non-diffusive contributions to the MC metric, we infer how 
long a simulation must be to be considered effectively ergodic. Second, once 
the ergodic limit is reached, we 
can compare the results from different numerical algorithms 
to measure relative efficiencies. Because the outcomes of MC simulations are 
noisy, we have found it useful to separate diffusive and 
non-diffusive terms in the MC metric with a Fourier analysis so that we
can neglect
the high frequency components of the noise.
This technique has given reproducible
results.

To test the match between the stochastic model and actual Monte Carlo
simulations, we examine the approach to ergodic behavior in simulations of
Lennard-Jones clusters.  Recently
\cite{juan1,juan2}
we have studied the thermodynamic
properties of Lennard-Jones clusters as a function of temperature using
both J-walking 
\cite{frantz}
and parallel tempering methods.  
\cite{p1,p2,p3}
Both simulation
techniques require an initial high temperature that must be ergodic when
Metropolis Monte Carlo methods 
\cite{metropolis}
are used.  If the Metropolis method does
not give ergodic results at the initial high temperature, systematic
errors propagate to the lower temperatures in J-walking and parallel
tempering simulations, and the results can be flawed or meaningless.  In
most Monte Carlo simulations of clusters at finite temperatures, 
\cite{lee,labastie}
the
clusters are defined by enclosing the atoms within a constraining
potential about the center of mass of the system.  The constraining
potential is necessary because clusters at finite temperatures have
finite vapor pressures, and the association of any one atom with the
cluster can be ill-defined.  From experience 
\cite{juan1,juan2,juan3}
we have found that if the
radius of the constraining potential and the initial high temperature
are not both carefully chosen,
it can be difficult to
attain ergodicity with Metropolis methods.  A key concern then is the
choice of constraining radius and the choice of initial temperature.  We
verify the stochastic model by investigating Monte Carlo simulation
results as a function of the temperature and the size of the
constraining potential.

The contents of the remainder of this paper are as follows.  In Section
II we motivate the studies that follow by examining numerally the
behavior of a set of Monte Carlo simulations of a 13-particle
Lennard-Jones cluster.  This cluster system is used to illustrate the
results throughout this paper.
In Section
III we introduce the stochastic model based on a continuous time
sequence.  In Section IV we extend the model to discrete time sequences
characteristic of actual Monte Carlo simulations.  In Section V we test
the discrete stochastic model with applications to Lennard-Jones
clusters and in Section VI we summarize our conclusions.  Many of the key
derivations needed for the developments are found in two appendices.

\section{An Example Calculation}

Before discussing the major developments of this work, it is useful to
understand the nature of the problem we are attempting to solve by
examining some numerical results on a prototypical system.  We take the
13-particle Lennard-Jones cluster defined by the potential function

\begin{equation}
V({\bf x}) = 4\varepsilon 
\sum_{i=2}^N\sum_{j=1}^{i-1}\left[\left(\frac{\sigma}{r_{ij}}\right)^{12} - 
\left(\frac{\sigma}{r_{ij}}\right)^{6}\right]
+ \sum_{i=1}^NV_C( \vec{x}_i,R_c)
\label{pot}
\end{equation}

\noindent where $\varepsilon$ and $\sigma$ 
are the standard Lennard-Jones energy and length parameters,
$N$ is the number of particles in the cluster (13 in the present case), 
$r_{ij}$ is the distance 
between particles $i$ and $j$

\begin{equation}
\label{rij}
r_{ij} = |\vec{x}_i-\vec{x}_j|,
\end{equation}

\noindent and $V_C$ is the constraining potential discussed in Sec. I

\begin{equation}
\label{cp}
V_C(\vec{x}_i,R_c) = \left\{
\begin{array}{ll}
0&\left|\vec{x}_i-\vec{X}_c\right| < R_c\\
\infty&R_c<\left|\vec{x}_i-\vec{X}_c\right|  
\end{array}
\right.\,\,,
\end{equation}

\noindent where $\vec{X}_c$ is the coordinate of the 
center of mass of the cluster and
$R_c$ is the radius of the constraining sphere. 
The 13-particle Lennard-Jones cluster has a complex
potential surface with many minima separated by significant energy
barriers, 
\cite{freeman1,wales1,wales2}
and ergodicity problems associated with the simulation of
properties of this system are well-known.
\cite{frantz}
We now consider a Metropolis
MC simulation of the average potential energy of the system in the
canonical ensemble at temperature $k_BT/\varepsilon=0.393$($k_B$ is the 
Boltzmann constant).  This average
potential energy $\overline{V}_k$ is defined by
\begin{equation}
\overline{V}_k=\frac{1}{k} \sum_{k'=1}^k V_{k'}
\end{equation}

\noindent and is displayed in the upper panel of Fig. 1 as a function
of the walk length $k$ for 20 independent simulations each
initialized from a random configuration.  Over the maximum time scale 
$K$ of the
walks, it apparent that the potential energy averaged over each
independent walk has not converged to the same result.  Such
unreproducible behavior is symptomatic of a simulation not yet at the
ergodic limit.

\begin{figure}
\epsfxsize=.48\textwidth
\epsfbox{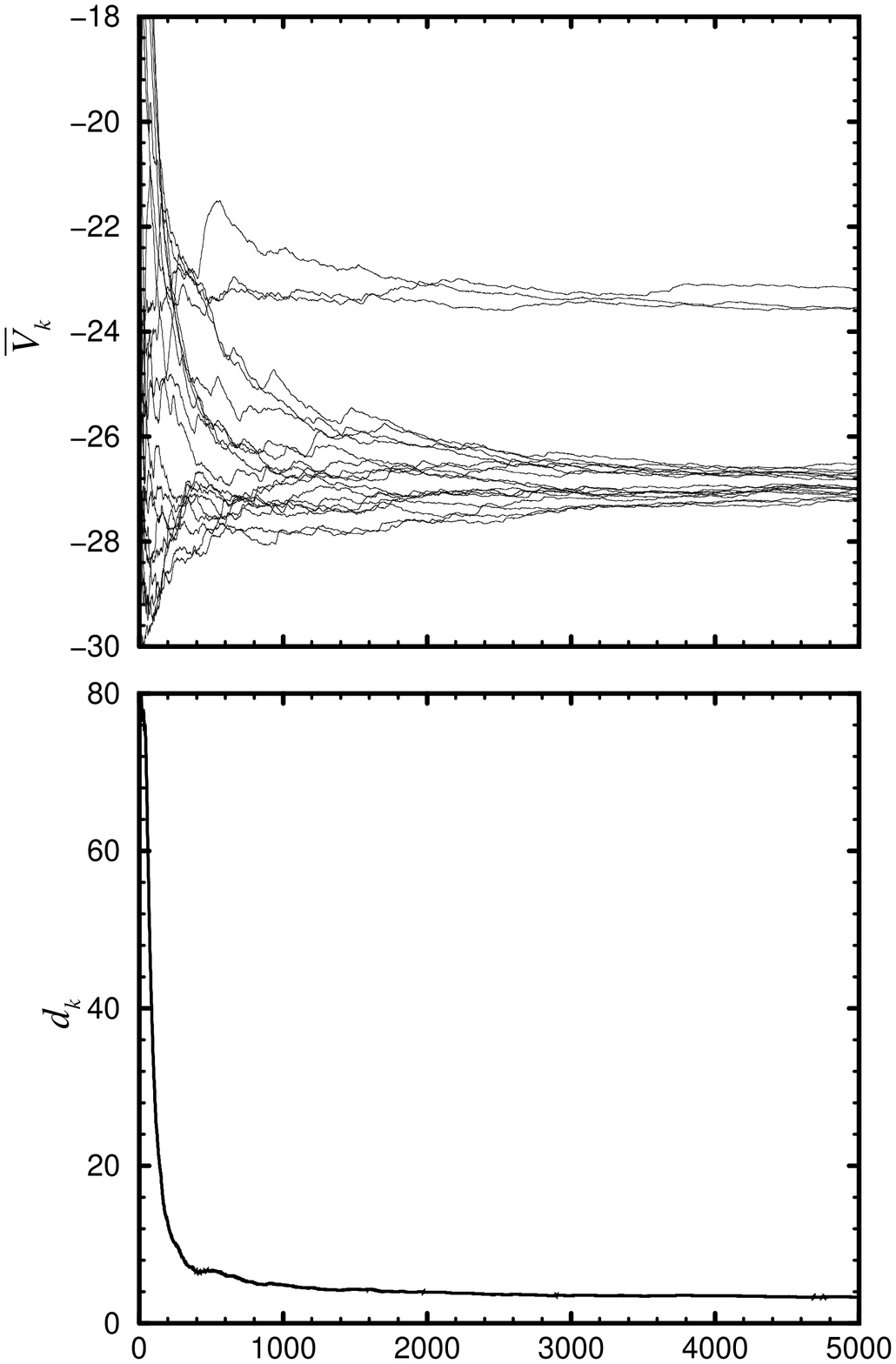}
\end{figure}

{\footnotesize
{\bf Fig. 1:} The upper panel shows the ``time evolution" of 
$\overline{V}_k$ (in units of $\varepsilon$) 
for $M=20$ independent experiments. The lower panel 
shows $d_k$ (in units of $\varepsilon^2$) vs. $k$ for the experiments of the 
upper panel. 
$R_c$ has been set to $4\sigma$ and  $k_BT/\varepsilon=0.393$. At least 
two basins with different energies are present. Clearly, 
$d_k$ goes to a constant when $k$ is increased within the total time
scale of the simulation.}
\vspace{.5cm}

At the ergodic limit (i.e. for the maximum walk length $K$ greater than
that included in Fig. 1) the averages displayed in the upper panel of Fig. 1
must approach the same value for each walker.  Using related ideas
developed elsewhere,
\cite{thiru1,thiru2,thiru3}
the extent to which the walks approach the same limit can be measured in
terms of a metric $d_k$ defined by

\begin{equation}
\label{omega1}
d_k = \frac2{M(M-1)} \sum_{i=2}^M \sum_{j=1}^{i-1} 
\left[\,\overline{V}^{(i)}_k - \overline{V}^{(j)}_k\right]^2\,\,,
\end{equation}

\noindent In Eq. (\ref{omega1}) $M$ represents the number of independent walks, 
and
$\overline{V}^{(i)}_k$ is the 
average potential energy computed in walk $i$ at MC 
time $k$.  The metric measures the energy fluctuations in the walk as a
function of the walk length.  For an ergodic simulation, the metric must
decay to zero.
For the 20 simulations of the 13-particle
Lennard-Jones cluster, the metric as a function of $k$ is plotted in the
lower panel of Fig. 1.  Rather than asymptotically approaching zero, over
the short length of the walk displayed here, $d_k$ has decayed to a
constant, and as discussed later in this paper, over the time scale of
this simulation, $d_k$ can be qualitatively represented by the function

\begin{equation}
d_k=\frac{A_K}{k}+B_K
\end{equation}

\noindent where $A_K$ and $B_K$ are coefficients that are dependent
on the total walk
length $K$. As $K$ is increased to a time
where the walk is ergodic,
$B_K$ must decay to zero.  Major goals of this work are to understand
how $B_K$ decays and to use the discovered decay law to determine the
onset of ergodic behavior.
Our approach is to introduce first a continuous
stochastic model of a simulation followed by a discrete model more
clearly linked to actual MC studies.

\section{Stochastic Model}

We have discussed in the introduction how the output of MC simulations can be 
considered to be 
a combination of stochastic processes with different time scales, 
and how the contributions to autocorrelation function from these
processes can vary when the length of the 
simulation is enlarged. Here we present a continuous time model for the 
stochastic processes that occur in a simulation. Even though a MC simulation 
occurs 
in a discrete time (each MC point represents a time unit), we find that the 
continuous model helps to understand better the ideas used in the 
modeling of the MC output.

In this section the ensemble mean value is used to find the expression 
for the autocorrelation functions of the model. Although in actual numerical 
calculations the ensemble mean is replaced by a mean over a finite number of 
independent experiments, the results obtained here give information about the 
limit of an infinite sample.

The stationary process used to sample space is a stochastic 
process. We assume the output of the MC simulation can be modeled by a linear 
superposition of stochastic processes with different 
correlation times  $\tau_{\ell}\ge 0$, 

\begin{equation}
\label{eq:2}
U(t) = U_c +  \sqrt{\Gamma_0} \,\xi(t)
        + \sum_{\ell=1}^{\Lambda} 
\sqrt{\Gamma_{\ell}}\,g_{\ell}(t/\tau_{\ell})\,\,,
\end{equation}

\noindent where $U_c$ is a constant,
the random variable $\xi(t)$ represents
white noise processes with
zero correlation time ($\tau_0=0$), and the $\{g_{\ell}(t/\tau_{\ell})\}$ 
are stochastic processes 
with correlation times $\tau_{\ell}>0$. 
$\xi(t)$ and $g_{\ell}(t/\tau_{\ell})$ 
have 
units of the square root of time, and $\Gamma_0$ and $\Gamma_{\ell}$ are constants
with units of $U^2/t$. If $U$ is chosen to be the the $x$-coordinate
of a particle, $\Gamma_0$ and $\Gamma_{\ell}$ have units of a
diffusion constant.  Consequently we refer to these constants as
generalized diffusion coefficients.  The white noise process has the following 
properties \cite{gardiner}

\begin{eqnarray}
\langle \xi(t) \rangle &=& 0 \label{noise1}\\
\langle \xi(t)\, \xi(t') \rangle &=& \delta(t-t')\label{noise2}\,\,,
\end{eqnarray}

\noindent and the remaining colored noise processes are assumed to satisfy

\begin{eqnarray}
\langle g_{\ell}(t/\tau_{\ell}) \rangle &=& 0 \label{g1}\\
\langle g_{\ell}(t/\tau_{\ell})\, g_{\ell}(t'/\tau_{\ell}) 
\rangle &=&  \frac1{\tau_{\ell}}\label{g2}\,f_{\ell}
\left(\frac{|t-t'|}{\tau_{\ell}}\right)
\label{ggg}
\end{eqnarray}

\noindent so that they represent processes with a memory $f_{\ell}$. 
Even though correlations between processes with different 
correlation times may be non-zero, 
we assume the processes to be independent, i.e.

\begin{eqnarray}
\langle g_{\ell}(t/\tau_{\ell})\,g_{\ell'}(t'/\tau_{\ell'})\rangle
&=&\langle g_{\ell}(t/\tau_{\ell})\rangle\langle g_{\ell'}(t'/\tau_{\ell'})
\rangle\nonumber\\
&=&0\,\,\,\,\forall\,\ell\neq\ell'
\label{a1}\\
\langle \xi(t)\, g_{\ell}(t'/\tau_{\ell})\rangle&=&\langle \xi(t)
\rangle\langle  g_{\ell}(t'/\tau_{\ell})\rangle\nonumber\\
&=&0\,\,\,\,\forall\,\,t\,{\rm and}\,t'
\label{a2}\,\,.
\end{eqnarray}

\noindent The memory function is assumed to be a continuous function 
that depends only on the distance between $t$ and $t'$ disregarding 
the time origin (stationary condition). The memory function represents 
the correlation between two times of the process $g_{\ell}$. In our model 
we impose the condition

\begin{equation}
\frac{t}{\tau_{\ell}}f_{\ell}\left( \frac{t}{\tau_{\ell}}\right)
<\int_0^t\!\!dt'\,\,\frac1{\tau_{\ell}}f_{\ell}\left( \frac{t'}
{\tau_{\ell}}\right)<\infty\label{F}\,\,.
\end{equation}

\noindent The scope and implications
of the leftmost inequality are explored in Appendix A. In 
Appendix A we also examine the conditions $f_{\ell}$ must satisfy in order to 
yield contributions to autocorrelation function that decay more weakly
than $1/t$.
We now assume that this inequality can be taken as a bound to possible 
maxima of $f_{\ell}$ appearing at $t>0$. The rightmost inequality enables 
us to assume $f_{\ell}$ is normalized

\begin{equation}
\label{norm}
\int_{-\infty}^{\infty}\!\!dt\,\,\frac1{\tau_{\ell}}f_{\ell}
\left( \frac{|t|}{\tau_{\ell}}\right)=1\,\,.
\end{equation}

\noindent We have identified here the time scale $\tau_{\ell}$ with the 
correlation time of the stochastic process $g_{\ell}$. This identification is 
valid if

\begin{equation}
\int_{-\infty}^{\infty}\!\!dt\,\,\frac{|t|}{\tau_{\ell}}f_{\ell}
\left( \frac{|t|}{\tau_{\ell}}\right)= \tau_{\ell},
\end{equation}

\noindent which implies that the behavior of $f_{\ell}$ at large $t$ must be
${\cal O}\left(t^{-(2+\epsilon)}\right)$, or smaller.

In addition, by the properties of the ensemble mean value, 
we have that for all real $\lambda$

\begin{eqnarray}
0&\leq& \left\langle \left[ g_{\ell}(t/\tau_{\ell})\,+\,\lambda\, g_{\ell}(t'/\tau_{\ell})\right]^2\right\rangle \nonumber\\
&\leq& 
\left\langle g_{\ell}(t/\tau_{\ell})^2\right\rangle + 2\lambda\,\left\langle g_{\ell}(t/\tau_{\ell})\, g_{\ell}(t'/\tau_{\ell}) \right\rangle
+ \lambda^2\left\langle g_{\ell}(t'/\tau_{\ell})^2\right\rangle
\nonumber\\
&\leq& \frac1{\tau_{\ell}}\left\{ f_{\ell}(0) + 2\lambda\, f_{\ell}\left(\frac{|t-t'|}{\tau_{\ell}}\right) + \lambda^2\, f_{\ell}(0) \right\}
\label{a3}\,\,.
\end{eqnarray}

\noindent Equation (\ref{a3}) must be true for all $\lambda$. 
Therefore, the discriminant of the polynomial in $\lambda$ must be 
non-positive

\begin{equation}
\label{schwartz}
4\,\left[f_{\ell}\left(\frac{|t-t'|}{\tau_{\ell}}\right) - 
\, f_{\ell}(0)\right]
\left[f_{\ell}\left(\frac{|t-t'|}{\tau_{\ell}}\right) + 
\, f_{\ell}(0)\right]\leq0\,\,.
\end{equation}

\noindent Consequently, $f_{\ell}(0)=\max\{f_{\ell}(x)\,\forall\,x\geq0\}$. 
Other properties of $f_{\ell}$ are studied in Appendix A.

The ensemble mean value $\langle U\rangle$ is time independent. The ensemble
mean value of
the noise processes is zero. Therefore, $U_c$ must be equal
to $\langle U\rangle$.
Processes defined by Eq. (\ref{eq:2}) have 
two different components, uncorrelated white noise and correlated processes 
with correlation time $\tau_{\ell}$.
Because the goal of the simulation is the calculation of the ensemble 
mean $\langle U \rangle$ by the analysis of the time series, we study the 
behavior of the temporal mean $\overline{U}(t)$

\begin{eqnarray}
\overline{U}(t) &=& \frac1t \int_0^t\!\!dt'\,\,U(t')\nonumber\\
&=& \langle U\rangle + \frac1t \sqrt{\Gamma_0} \,W(t) +
\frac1t \sum_{\ell=1}^{\Lambda} 
\sqrt{\Gamma_{\ell}}\,G_{\ell}(t/\tau_{\ell})\,\,,
\label{eq:3}
\end{eqnarray}

\noindent where $W(t)$ is a Wiener process, \cite{gardiner}

\begin{eqnarray}
W(t)&=&\int_0^t\!\!dt'\,\,\xi(t')\\
\langle W(t)\rangle &=& 0 \label{wiener1}\\
\langle W(t)\,W(t')\rangle &=& t_<\,\,,\label{wiener2}
\end{eqnarray}

\noindent with $t_<=\min(t,t')$, and 
$G_{\ell}(t/\tau_{\ell})=\int_0^t\!dt'\,\, g_{\ell}(t'/\tau_{\ell})$. 

Equation (\ref{eq:3}) implies that the evolution of the temporal 
mean $\overline{U}(t)$ has the same structure as $U$, with an uncorrelated 
term and terms with tailed correlation functions. 

The autocorrelation function of the process $\overline{U}$ at 
times $t$ and $t'$ is defined by

\end{multicols}

\begin{eqnarray}
\kappa(t,t')&=&\left\langle\left(\overline{U}(t)-\langle U\rangle\right)
\left(\overline{U}(t')-\langle U\rangle\right)
\right\rangle \nonumber\\
&=& \frac{\Gamma_0}{tt'}\,\left\langle W(t)\,W(t')\right\rangle +
\frac1{tt'}\sum_{\ell=1}^{\Lambda}\Gamma_{\ell}\,\left\langle 
G_{\ell}(t/\tau_{\ell})\,G_{\ell}(t'/\tau_{\ell})\right\rangle\,\,,
\label{c1}
\end{eqnarray}

\noindent where we have used Eqs. (\ref{a1}) and (\ref{a2}) to 
neglect terms involving processes with different correlation times.

Because we have assumed $f_{\ell}$ is a continuous function, 
$f_{\ell}$ reaches its maximum and minimum value within any closed 
interval considered. The $\ell$th non-diffusive contribution 
to $\kappa(t,t')$
 
\begin{equation}
\frac1{tt'}\,\langle G_{\ell}(t/\tau_{\ell})\,G_{\ell}
(t'/\tau_{\ell})\rangle =\frac1{tt'}\int_0^t\!\! dt_1
\int_0^{t'}\!\! dt_2 \,\,\frac1{\tau_{\ell}}f_{\ell}
\left(\frac{|t_1-t_2|}{\tau_{\ell}}\right)\,\,,
\end{equation}

\noindent is bounded

\begin{eqnarray}
\frac1{t_<t_>}\int_0^{t_<}\!\!dt_1\,\,\int_0^{t_>}\!\!dt_2\,
\,\frac1{\tau_{\ell}}f_{\ell}\left( \frac{t_{min}}{\tau_{\ell}}\right)
\leq \frac1{tt'}&\langle G_{\ell}(t/\tau_{\ell})\,G_{\ell}(t'/\tau_{\ell})
\rangle\;&\leq 
\frac1{t_<t_>}\int_0^{t_<}\!\!dt_1\,\,\int_0^{t_>}\!\!dt_2\,\,
\frac1{\tau_{\ell}}f_{\ell}(0)
\nonumber\\
\frac1{\tau_{\ell}}f_{\ell}\left(\frac{t_{min}}{\tau_{\ell}}\right)
\leq
\frac1{tt'}&\langle G_{\ell}(t/\tau_{\ell})\,G_{\ell}(t'/\tau_{\ell})
\rangle\;&\leq 
\frac1{\tau_{\ell}}f_{\ell}(0)\label{bb}\,\,,
\end{eqnarray}

\begin{multicols}{2}

\noindent where $t_>=\max(t,t')$, and $t_{min}$ is the time at which 
$f_{\ell}$ reaches its minimum value in the closed interval $[0,t_>]$. 
There exists a  $t^*_{\ell}(t_>)\in[0,t_{min}]$ \cite{spivak} such that,

\begin{equation}
\frac1{tt'}\,\langle G_{\ell}(t/\tau_{\ell})\,G_{\ell}(t'/\tau_{\ell})\rangle 
= \frac1{\tau_{\ell}}f_{\ell}\left(\frac{t_{\ell}^*(t_>)}{\tau_{\ell}}
\right)\,\,.
\label{tl}
\end{equation}

\noindent
Using Eqs. (\ref{wiener2}) and (\ref{tl}) in (\ref{c1}), we find that

\begin{equation}
\label{correl}
\kappa(t,t')= \frac{\Gamma_0}{t_>} + \sum_{\ell=1}^{\Lambda}
\frac{\Gamma_{\ell}}{\tau_{\ell}} \, 
f_{\ell}\left(\frac{t_{\ell}^*(t_>)}{\tau_{\ell}}\right)\,\,.
\end{equation}

\noindent For all times shorter than $\tau_1$ the autocorrelation function 
is the sum of {\em diffusive contributions} (proportional to $1/t$) plus 
{\em non-diffusive contributions}. These contributions implicitly depend on 
$t_>$ through $t^*_{\ell}(t_>)$. We assume that $f_{\ell}$ satisfies the 
conditions stated in Appendix A, so that the
dependence of $f_{\ell}$ on $t$ is {\em weaker than} $1/t$ 
(for total time scales shorter than $\tau_{\ell}$;  see Appendix A).

We next consider the behavior of Eq. (\ref{correl}) for time scales
greater than $\tau_1$.
Under the scale change $t \to bt$ such that $\tau_1\ll bt_{>} \ll \tau_2$, 
the contributions to the correlation function from the process with 
correlation time $\tau_1$ can be considered diffusive [in other words, 
by virtue of Eqs. (\ref{noise2}) and (\ref{ggg}), $f_1/\tau_1$ has become a 
delta function]. With $bt_{>} \ll \tau_2$, the other processes preserve 
their old properties. Then, the autocorrelation function can be expressed

\begin{equation}
\label{urenor}
\kappa(bt,bt') =
\frac{\Gamma_0+\Gamma_1}{bt_>}+\sum_{\ell=2}^{\Lambda}
\frac{\Gamma_{\ell}}{\tau_{\ell}} \, f_{\ell}
\left(\frac{t_{b\ell}^*(t_>)}{\tau_{b\ell}}\right)\,\,.
\end{equation}

\noindent The complete derivation of Eq. (\ref{urenor}) can be found in 
Appendix B.
For a times larger than the correlation time $\tau_{\Lambda}$, all contributions
to the autocorrelation function are diffusive,
the simulation can be considered ergodic, the sampling complete, 
and the temporal mean is equal to the ensemble mean within ${\cal O}(1/t)$ 
mean square fluctuations.

\section{Discrete time sequences and the MC metric}

Monte Carlo simulations generate discrete sequences $U_k$ of values of the 
quantity under study. Additionally,
in actual calculations the ensemble of sequences is represented by a
finite rather than an infinite set.  In this section, the model developed in 
the previous section is extended to finite sets of discrete sequences. 
We express the
$M$ sequences $\,\left\{U_k^{(m)}\right\}_{k=1}^K\,$, where the label $(m)$ 
ranges
from 1 to $M$. The exact ensemble mean value $\langle U\rangle$ can 
be obtained
in the limit that $M$ becomes infinite.  In analogy with the model developed 
in Section III, each output is assumed to have the form

\begin{equation}
\label{seq}
U_k^{(m)} = \langle U\rangle + \sqrt{\Gamma_0}\,\xi_k^{(m)} + \sum_{\ell=1}^
{\Lambda}\sqrt{\Gamma_{\ell}}\,g_{\ell;\,k/\tau_{\ell}}^{(m)}\,\,,
\end{equation}

\noindent where

\begin{eqnarray}
\langle \xi_k^{(m)}\rangle&=&0\\
\langle \xi_k^{(m)}\,\,\xi_{k'}^{(n)}\rangle&=&\delta_{m,n}\,\delta_{k,k'}\\
\langle g_{\ell;\,k/\tau_{\ell}}^{(m)}\rangle&=&0\\
\langle g_{\ell;\,k/\tau_{\ell}}^{(m)}\,\,
g_{\ell';\,k'/\tau_{\ell'}}^{(n)}\rangle&=&
\delta_{m,n}\,\delta_{\ell,\ell'}\,f_{\ell}\left(\frac{|k-k'|}{\tau_{\ell}}
\right)\,\,.
\end{eqnarray}

\noindent The true ensemble
average $\langle U\rangle$
does not depend on the index $m$. 

In the discrete case we define a metric

\begin{equation}
\label{omega}
d_k = \frac2{M(M-1)} \sum_{i=2}^M \sum_{j=1}^{i-1} 
\left[\,\overline{U}^{(i)}_k - \overline{U}^{(j)}_k\right]^2\,\,,
\end{equation}

\noindent where the bars represent the temporal mean value

\begin{eqnarray}
\overline{U}^{(m)}_k &=& \frac1k \sum_{k'=1}^k U^{(m)}_{k'}\nonumber\\
&=&\langle U\rangle + \frac{\sqrt{\Gamma_0}}k\,W_k^{(m)}+ \sum_{\ell=1}^{\Lambda}\frac{\sqrt{\Gamma_{\ell}}}k\,G_{\ell;\,k/\tau_{\ell}}
^{(m)} \,\,,
\end{eqnarray}

\noindent with

\begin{eqnarray}
W_k^{(m)} &=& \sum_{k'=1}^k \xi_{k'}^{(m)}\\
G_{\ell;\,k/\tau_{\ell}}^{(m)} &=& 
\sum_{k'=1}^k g_{\ell;\,k'/\tau_{\ell}}^{(m)}\,\,.
\end{eqnarray}

\noindent Observe that in the present case, our finite sample 
of the infinite ensemble is the set of outcomes from $M$ independent 
numerical experiments. The metric we have defined in Eq. (\ref{omega})
can be contrasted with alternative metrics 
\cite{thiru1,thiru2,thiru3}
previously defined for molecular dynamics simulations.
These alternative metrics examine
the fluctuations of two simulations initialized from different components
of configuration space averaged with respect to all the
particles in the system.  The metric we use in this work is determined
using an
average with respect to $M$ independent 
simulations that represent a subset of the full
ensemble.

Using the model presented in Eq. (\ref{seq}), we now develop 
a way
to predict the behavior of the MC simulation in the 
non-ergodic and the ergodic regimes.
We first consider the case that the total simulation time 
$K$ is larger than the first 
correlation time $\tau_1$ but shorter than $\tau_2$, i.e. 
$\tau_1\ll K\ll\tau_2$. The expression for 
$d_k$ is given by

\end{multicols}

\begin{eqnarray}
d_k &=& \frac2{M(M-1)} \sum_{i=2}^M \sum_{j=1}^{i-1} 
\left[ \left(\overline{U}^{(i)}_k-\langle U\rangle\right) - 
\left(\overline{U}^{(j)}_k-\langle U\rangle\right)\right]^2\nonumber\\
&=& \frac2M \sum_{i=1}^M\left(\overline{U}^{(i)}_k-\langle U\rangle\right)^2 - 
\frac4{M(M-1)} \sum_{i=2}^M \sum_{j=1}^{i-1} 
\left(\overline{U}^{(i)}_k-\langle U\rangle\right) \, 
\left(\overline{U}^{(j)}_k-\langle U\rangle\right)\nonumber\\
&=& 2\,\Gamma_0 \,\frac1M \sum_{i=1}^M
\left(\frac{W_k^{(i)}}k\right)^2 + 
2\,\Gamma_1 \,\frac1M \sum_{i=1}^M
\left(\frac{G_{1;\,k/\tau_1}^{(i)}}k\right)^2 +
2\,\sum_{\ell=2}^{\Lambda}\Gamma_{\ell} \,\frac1M \sum_{i=1}^M
\left(\frac{G_{\ell;\,k/\tau_{\ell}}^{(i)}}k\right)^2 \nonumber\\
&&+ \,4\,\sum_{\ell=1}^{\Lambda}\sqrt{\Gamma_0 
\Gamma_{\ell}} \,\,\frac1M \sum_{i=1}^M\frac{W_k^{(i)}
\,G_{\ell;\,k/\tau_{\ell}} ^{(i)}}{k^2}
 + \,4\,\sum_{\ell=2}^{\Lambda}\sum_{\ell'=1}^{\ell-1}\sqrt{\Gamma_{\ell} 
\Gamma_{\ell'}} \,\,\frac1M \sum_{i=1}^M\frac{G_{\ell;\,k/\tau_{\ell}}^{(i)}
\,G_{\ell';\,k/\tau_{\ell'}} ^{(i)}}{k^2} \nonumber\\
&&- \frac4{M(M-1)} \sum_{i=2}^M 
\sum_{j=1}^{i-1} \left(\overline{U}^{(i)}_k-\langle U\rangle\right) \, 
\left(\overline{U}^{(j)}_k-\langle U\rangle\right)\,\,.  
\end{eqnarray}

\noindent If the number of experiments $M$ is sufficiently large, 
we can neglect terms involving processes with different correlation times, 
and products of sequences belonging to different experiments. 
Under these assumptions we obtain

\begin{equation}
\label{om}
d_k = 2\,\frac{\Gamma_0}k \,\frac1M \sum_{i=1}^M
\frac{{W_k^{(i)}}^2}k +
2\,\frac{\Gamma_1}k \,\frac1M \sum_{i=1}^M
\frac{{G_{1;\,k/\tau_1}^{(i)}}^2}k +
    2\,\sum_{\ell=2}^{\Lambda}\Gamma_{\ell} \,\frac1M \sum_{i=1}^M
\left(\frac{G_{\ell;\,k/\tau_{\ell}}^{(i)}}k\right)^2\,.
\end{equation}

\begin{multicols}{2}

\noindent Equation (\ref{om}) preserves the form of Eq. (\ref{correl}). 
To make this statement explicit, let us rewrite Eq. (\ref{om}) as

\begin{equation}
d_k = 2\,\frac{\Gamma_k}k + 2\,\Upsilon_k\,\,,
\end{equation}

\noindent where

\begin{eqnarray}
\Gamma_k   &=& \Gamma_0\,\frac1M \sum_{i=1}^M
\frac{{W_k^{(i)}}^2}k +
\Gamma_1\,\frac1M \sum_{i=1}^M
\frac{{G_{1;\,k/\tau_1}^{(i)}}^2}k\label{g} \label{eq:gmmg} \\
\Upsilon_k &=& \sum_{\ell=2}^{\Lambda}\Gamma_{\ell} \,\frac1M \sum_{i=1}^M
\left(\frac{G_{\ell;\,k/\tau_{\ell}}^{(i)}}k\right)^2\,\,.
\label{ups}
\end{eqnarray}

\noindent In Appendix B we present a study of the way non-diffusive contribution become diffusive under time scale 
changes. If $M$ is sufficiently large and $\tau_1\ll K \ll \tau_2$, 
by virtue of Appendix 
B, $\Gamma_k$ must roughly be a constant.
By {\em roughly a constant} we mean a 
constant $C$ plus some rapidly fluctuating function $\zeta_k$, with the 
following properties: a) $\langle\zeta_k\rangle=0$ and b) 
$|C|\gg\max_{k=1,2,\dots,K}(|\zeta_k|)$. Then

\begin{equation}
\label{gz}
\Gamma_k \simeq \Gamma_K + \zeta_k
\end{equation}
 
\noindent If $K$ is enlarged, we expect to 
have a larger 
value of $\Gamma_K$. 
$\Upsilon_k$ is a quantity related to the memory functions 
$f_{\ell}$ with 
correlation times $\tau_{\ell}\gg K$. In the continuous 
time model, the colored noise 
processes contribute to the autocorrelation function 
with terms proportional to $f_{\ell}(t^*_{\ell}(t_>)/\tau_{\ell})$, which are 
weakly dependent on $t$ (see Appendix A). 
We can expect $\Upsilon_k$ to be
weakly dependent on $k$, and for sequences of length $K$ and for $M$ 
sufficiently large, we consider this quantity roughly to be a constant

\begin{equation}
\label{uz} 
\Upsilon_k\simeq\Upsilon_K+\beta_k\,\,.
\end{equation}

\noindent where $\beta_k$ represents additional random noise.
Then, for a given length 
$k\leq K$, the MC metric $d_k$ can be approximated by 

\begin{equation}
\label{d}
d_k = 2\,\frac{\Gamma_K}k + 2\,\Upsilon_K+\gamma_k\,\,,
\end{equation}

\noindent where $\gamma_k=2(\zeta_k/k+\beta_k)$ represents remaining stochastic noise 
from
both contributions.  In  
this approximation, $\Gamma_K$ and $\Upsilon_K$ are the quantities 
that 
carry the long time dependence. Short time features appear in the $1/k$ 
dependence and in the remaining noise $\gamma_k$. If the sequences considered 
are increased in size by a factor of $b$, such that 
$\tau_{\lambda-1}\ll K\ll\tau_{\lambda}\ll bK$ for a given 
$1\leq\lambda\leq\Lambda$, $\Gamma_K$ ($\Upsilon_K$) is 
increased (decreased) (see Appendix B). Then,

\begin{equation}
\label{di}
d_{bk} = 2\,\frac{\Gamma_{bK}}{bk} + 2\,\Upsilon_{bK}+\gamma_{bk}\,\,,
\end{equation}

\noindent where $\Upsilon_{bK}$ must go to zero and 
$\Gamma_{bK}$ must approach a constant when $b$ is increased. 
By virtue of the expected behavior of the non-diffusive contributions (see 
Appendix A), we propose the following expression for $\Upsilon_{bK}$

\begin{equation}
\label{upa}
\Upsilon_{bK}=\Upsilon_K\,\phi(b)\,\,,
\end{equation}

\noindent where $\phi(b)$ is a decreasing function of $b$. Moreover, 
$\Upsilon_{bK}$ is a sum of non-diffusive contributions. As presented in 
Appendix A, each non-diffusive 
contribution to the autocorrelation function has a relative variation smaller 
than the relative variation of the diffusive contribution, namely $1-1/b$. If 
this inequality
is applicable to the sum of non-diffusive contributions, we have that

\begin{eqnarray}
1-\frac1b &>& 1-\frac{\Upsilon_{bK}}{\Upsilon_K}\\
1-\frac1b &>& 1-\phi(b)\\
1&<&b\,\phi(b)\label{bphi}\,\,,
\end{eqnarray}

\noindent for all $b>1$. Then, $\phi$ must be either

\begin{equation}
\phi(b)=b^{-\upsilon}\label{upa1}
\end{equation}

\noindent or

\begin{equation}
\phi(b)=\frac1{\eta\,\ln(b)+1}\label{upa2}\,\,,
\end{equation}

\noindent with $0<\upsilon<1$ and $0<\eta \le 1$. Equation (\ref{upa2}) can be 
thought as the limit
of Eq. (\ref{upa1}) when the exponent goes to zero. We know of no 
{\em a priori}
argument to 
justify Eq. (\ref{upa}). However, as is discussed in Section V, 
our numerical experience has shown Eq. (\ref{upa}) to be obeyed in all
cases we have examined.

Our goal is to develop a criterion to decide when the 
simulation can be 
considered ergodic. From the previous considerations it is clear that the 
ergodic limit is reached when $\Upsilon_K$ is indistinguishable from zero. 
The output from a MC simulation is usually noisy. Therefore, $\gamma_k$ 
can not 
 be neglected. A useful way to separate diffusive 
and non-diffusive 
contributions and to eliminate the stochastic noise from Eq. (\ref{d}), is to 
perform a Fourier analysis of the function $kd_k$. Let us define the 
frequencies $\omega_n=(2\pi/K)\,n$, with $n=0,1,\dots,K-1$. The discrete 
Fourier transform of the function $kd_k$ is the signal $Y_K(\omega_n)$

\end{multicols}

\begin{eqnarray}
Y_K(\omega_n)&=&\widehat{kd_k}(\omega_n)=\frac1K\sum_{k=1}^K 
\exp(-i\omega_nk)\,kd_k \label{eq:ft}\\
&=& \frac2K\sum_{k=1}^K \exp(-i\omega_nk) \,\,\Gamma_K\,+\,
\frac2K\sum_{k=1}^K k\,\exp(-i\omega_nk) \,\,\Upsilon_K+
\widehat{k\gamma_k}(\omega_n)\nonumber\\
&=& 2\,\delta_{n,0}\,\Gamma_K + \left\{\delta_{n,0} (K+1) + 
\left(1-\delta_{n,0}\right)\left(1+i\,\cot(\omega_n/2)\right)\right\}
\Upsilon_K
+\widehat{k\gamma_k}(\omega_n)\nonumber\\
&=& 2\,\delta_{n,0}\,\Gamma_K + \left(K\,\delta_{n,0}+1\right)\,
\Upsilon_K
+ i\,\left(1-\delta_{n,0}\right)\,\cot(\omega_n/2)\,\Upsilon_K+
\widehat{k\gamma_k}(\omega_n)
\label{fs}
\end{eqnarray}

\begin{multicols}{2}

\noindent In general, $\widehat{k\gamma_k}(\omega_n)$ 
is negligible except at high 
frequencies. For small positive values of the frequency we can make 
the approximation $\cot(\omega_n/2)\simeq 2/\omega_n$.  From this
approximation we have

\begin{equation}
\label{eq:10}
{\rm Im}\left(Y_K(\omega_n)\right)\simeq \frac2{\omega_n}\, 
\Upsilon_K\,\,.
\end{equation}

\noindent The real part of Eq. (\ref{fs}) for positive frequencies is

\begin{equation}
\label{ns}
{\rm Re}\left(Y_K(\omega_n)\right)=\Upsilon_K\,\,.
\end{equation}

\noindent Even though simpler than Eq. (\ref{eq:10}), we have
found Eq. (\ref{ns}) is more
sensitive to the deviations of $d_k$ from the approximation Eq. (\ref{d}).
Therefore, the data obtained from the real part is of poorer quality than
the data obtained from the imaginary part.

Equation (\ref{eq:10}) implies that for a given simulation 
length $K$, the contributions to the MC metric from the non-diffusive 
process can be determined from a simple relationship involving the 
Fourier transform of the function $kd_k$ at low frequencies. By increasing 
the length of the run $K$ by a factor of $b$, it is possible to observe the 
dependence of $\Upsilon_{bK}$ on $bK$. 

\section{Applications}

The concepts developed in the previous sections are sufficiently 
general to be applied to any kind of MC simulation.
We devote the present section to the application of the developments of
this paper
to the study of the Lennard-Jones 
13-particle cluster in the canonical
ensemble. This system has been introduced previously in Sec. II.

Some thermodynamic properties of clusters as a function of temperature
exhibit rapid changes that are reminiscent of similar changes that occur
for the same properties in bulk systems at phase transitions.  In a bulk
system a phase transition occurs at a single temperature.  For clusters
the rapid changes in thermodynamic properties occur over a finite
temperature interval.  To distinguish the temperature range where
thermodynamic properties change rapidly in clusters from a true phase
transition,  we follow Berry {\em et al.} 
\cite{berry}
and refer to such changes in
physical properties as associated with a phase change.  A common
property that has been found to be useful in monitoring these phase
change intervals of temperature is the heat capacity at constant 
volume \cite{clusters}

\begin{equation}
\label{cv}
C_V(T) = \frac1{k_BT^2}\,\left\langle(V-\langle V\rangle_T)^2
\right\rangle_T + \frac32Nk_B\,\,,
\end{equation}

\noindent where
$\langle\cdot\rangle_T$ represents the classical canonical mean value.

In this work we consider the bare Metropolis (Met), 
\cite{metropolis}
J-walking (Jw), \cite{frantz} and
parallel tempering (PT) \cite{p1,p2,p3} approaches to Monte Carlo simulations.
The free variable of all these methods is the reduced temperature 
$k_BT/\varepsilon$. In PT and Jw simulations, the highest temperature 
used ($T_h$) must be sufficiently large to ensure that Met is 
ergodic.\cite{frantz} 
From experience simulating a variety of systems, we have found
that $T_h$ must also be lower than a temperature $T_b$ where cluster
evaporation events become frequent.  It is useful to think of $T_b$ as
the cluster analogue of a boiling temperature.  We have found that Met
is unable to sample the boiling phase change region for clusters
ergodically, using total time scales accessible to current simulations.

For the results that follow, $U_k^{(m)}$ is chosen to be represented by
the potential energy of the system.  
In general $U_k^{(m)}$ can be any scalar property of the system.
We define a pass to
represent a set of single particle MC moves taken sequentially over the 13
particles in the cluster. We take $U_k^{(m)}$ to be the
potential energy at the $k'$th pass, in the $m'$th experiment. Using 
Eq. (\ref{fs}) we can write

\begin{equation}
\label{o1}
Y_K(0) = 2\,\Gamma_K+(K+1)\,\Upsilon_K\,\,.
\end{equation}

\noindent In the non-ergodic regime, $Y_K(0)$ grows 
with $K$, while in the ergodic regime, the signal $Y_K(0)$ 
approaches a constant. 

We begin by displaying results obtained for a calculation that has not
attained ergodicity over the time scale of the simulation.  We examine
the 13-particle Lennard-Jones cluster with the Met algorithm setting
$R_c=4\sigma$ at a temperature of $k_BT_h/\varepsilon=0.393$.  The
temperature is chosen to be that typically used as the initial high
temperature in Jw and PT studies of LJ$_{13}$.  By choosing a large
constraining radius, the evaporation events are so frequent at the
chosen temperature that attaining ergodicity proves to be quite
difficult.  We demonstrate the effect of reducing the constraining
radius shortly.  

\begin{figure}
\epsfxsize=.48\textwidth
\epsfbox{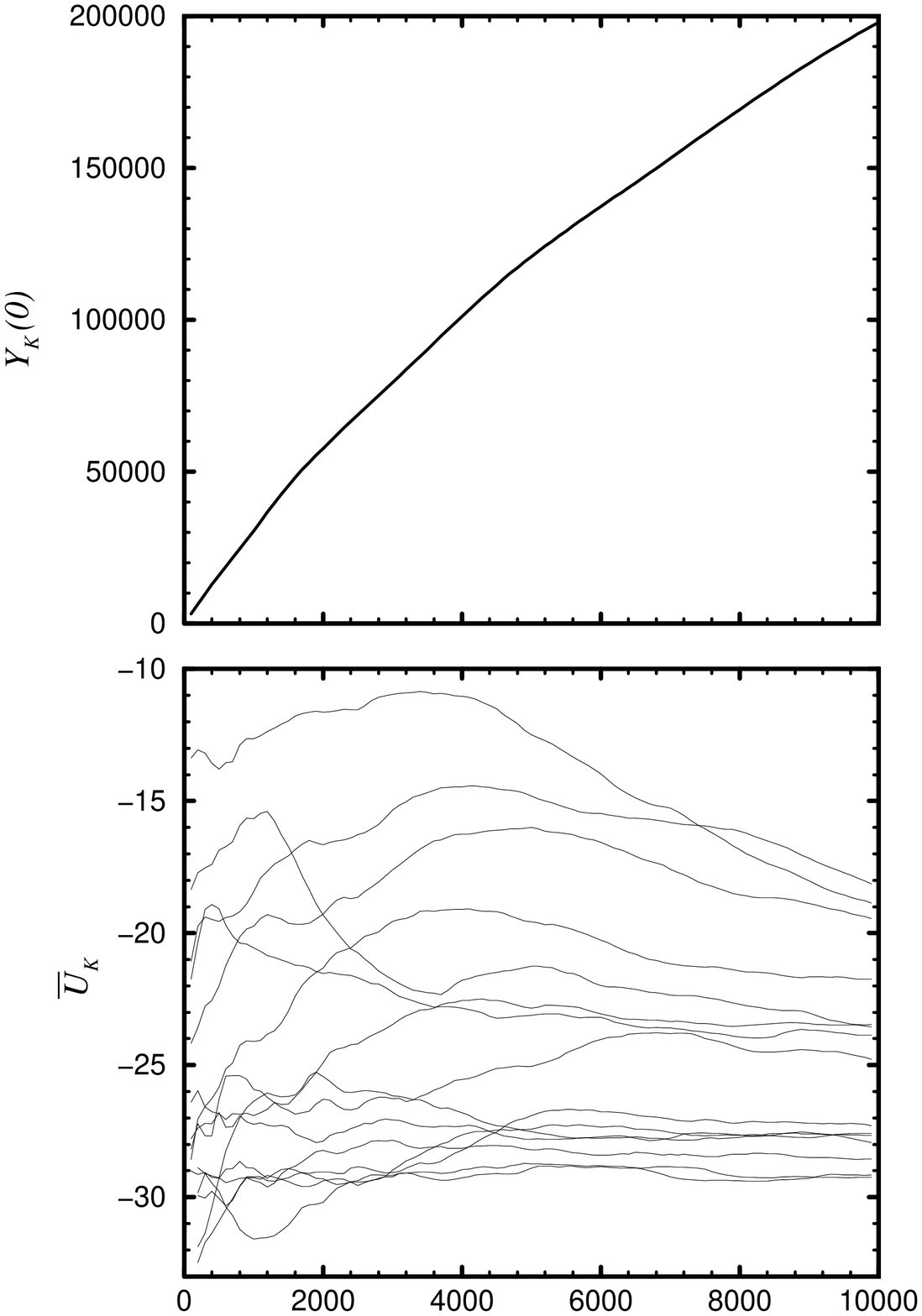}
\end{figure}

{\footnotesize
{\bf Fig. 2:} The upper panel is the signal $Y_K(0)$ (in units
of $\varepsilon^2$) vs. $K$ for 
$R_c=4\sigma$ at $k_BT_h/\varepsilon=0.393$. from $M=40$ independent 
experiments, of LJ$_{13}$. The length of the simulation is $10^4$ 
MC passes. The lower panel shows the ``time evolution" of $\overline{U}_K$ 
(in units of $\varepsilon$) for 15 independent experiments. At 
least three basins with different energies are present. Clearly, the 
simulation at this scale of time, is not ergodic.}
\vspace{.5cm}

\noindent The 
number of replicas used in the calculation is $M=40$, and $K=10^4$. 
The upper panel of Fig. 2
shows the signal $Y_K(0)$ [evaluated using Eq. (\ref{eq:ft})], 
which grows along the entire
simulation. This is the behavior expected in the non-ergodic regime. In the 
lower panel we can see 
the ``time evolution" of the temporal mean values of 15 experiments. 

\begin{figure}
\epsfxsize=.48\textwidth
\epsfbox{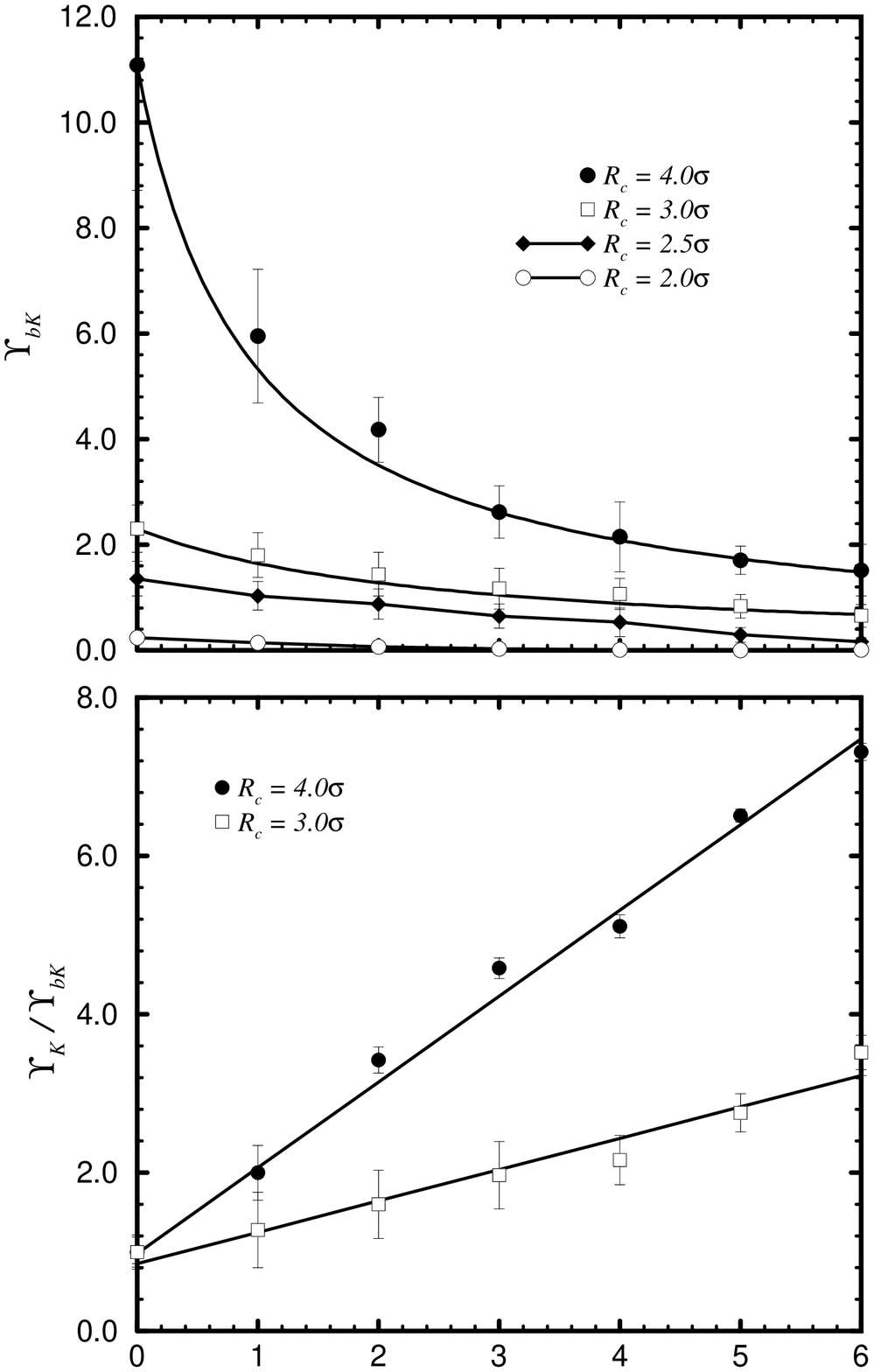}
\end{figure}

{\footnotesize
{\bf Fig. 3:} Upper panel: $\Upsilon_{bK}$ (in units
of $\varepsilon^2$) as a 
function of $\log_2(b)$ for $R_c=4\sigma$, $3\sigma$, $2.5\sigma$, and 
$2\sigma$. For the
two larger radii the full line is the best fit to the data points, 
according
to Eq. (\ref{upa}) with $\phi$ defined in Eq. (\ref{upa2}).
The lower panel shows the linear behavior of 
$\Upsilon_K/\Upsilon_{bK}$ vs. $\log_2(b)$, for 
$R_c=4\sigma$ and $3\sigma$. $K$ has been set to $10^4$.}
\vspace{.5cm}

\noindent There are three sets of curves, each of which is indicative of 
sampling of at least 
three different energy basins. At low values of $K$ the curves in the
lower panel differ significantly.
At $K\simeq4\,000$ the high energy basin curves begin to decrease
in energy. For a value of $K$ larger than the data displayed in Fig. 2,
the curves can be expected to coalesce with
the low energy basin curves. It is clear that for $K\leq10\,000$, the 
simulation is not ergodic.

In PT and Jw studies it is essential that the initial high temperature
walk be ergodic.  Ergodicity can be attained for LJ$_{13}$ by reducing
the radius of the constraining potential so that evaporation events are
rare.  We now present a study of $\Upsilon_K$ as a function of $K$ for
several values of $R_c$.
To determine $\Upsilon_K$, we have calculated the
Fourier transform function $Y_K(\omega_n)$ using Eq. (\ref{eq:ft}) 
at a
series of frequencies $\omega_n=2\pi n/K$ where $n$ has ranged from 1 to
$\min(\sqrt{12}bK/20\pi,100)$.  This range of frequencies ensures the linear
approximation used in Eq. (\ref{fs}) is valid while including sufficient
numbers of points for accuracy.\cite{citfreq}
Using Eq. (\ref{eq:10}), we have 
calculated the slope of the imaginary part of 
$1/Y_K(\omega_n)$ as a function of $\omega_n$, for 
these frequencies. The data points appearing in Fig. 3 are 
the mean value over twenty independent calculations of the slope of 
$1/Y_K(\omega_n)$.

\begin{figure}
\epsfxsize=.48\textwidth
\epsfbox{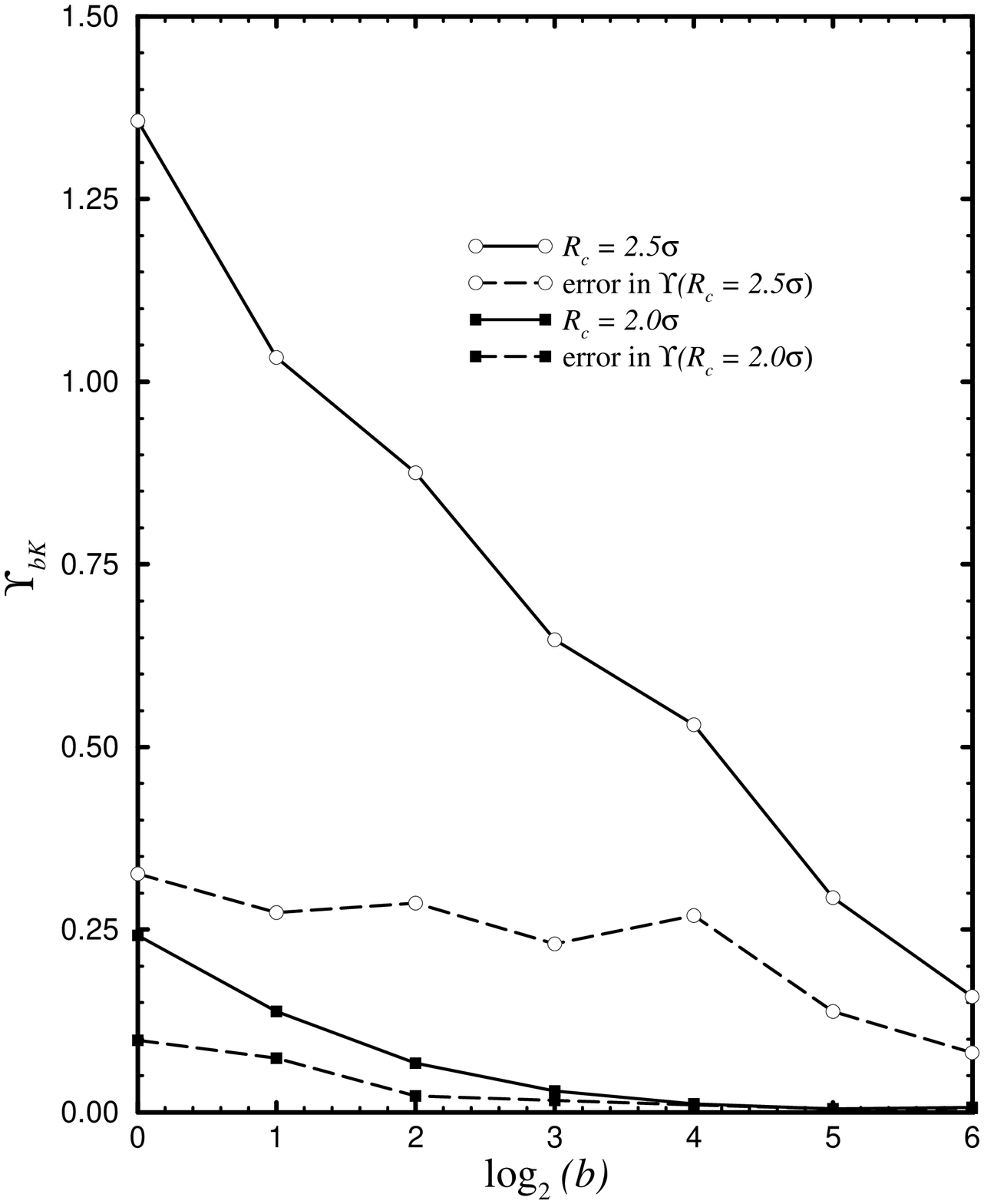}
\end{figure}

{\footnotesize
{\bf Fig. 4:} $\Upsilon_{bK}$ (in units
of $\varepsilon^2$) and its error vs. $\log_2(b)$ for 
$R_c=2.5\sigma$ and $2\sigma$, with $K=10^4$. When $\Upsilon_{bK}$ is on 
the order of its 
own error, the simulation can be 
considered ergodic. For $R_c = 2\sigma$ the simulation becomes ergodic at 
$\log_2(b)\simeq4$ ($bK\simeq16\times 10^4$). For $R_c=2.5\sigma$ a longer 
simulation is needed to reach ergodicity.}
\vspace{.5cm}

Starting from random configurations, we have performed $5\times10^4$ 
Met passes at $k_BT_h/\varepsilon=0.393$. After this warmup process, we 
have created sequences of sizes $bK=10^4$, $2\times10^4$, $4\times10^4$,
$\dots$, $64\times10^4$. 
The results are presented in Fig. 3 for $R_c=4\sigma$, $3\sigma$, 
$2.5\sigma$, and $2\sigma$.
The upper panel shows $\Upsilon_{bK}$ as a function 
of $\log_2(b)$, for fixed $K=10^4$. We have chosen to present the data
using base 2 logarithms for clarity (each increase by 1 unit of $\log_2(b)$
represents a factor of 2 scale increase).
All the data decrease with increasing
$b$, but only $R_c=2 \sigma$ and $R_c=2.5 \sigma$ appear to vanish
to within the error bars over the time scale of the current simulation.
In the lower panel we present
$\Upsilon_K/\Upsilon_{bK}$ as a function of
$\log_2(b)$ for $R_c=4$ and $3\sigma$. 
The decay law suggested in Eq. (\ref{upa}) with $\phi$ 
given by Eq. (\ref{upa2}) is satisfied for both radii.

\begin{figure}
\epsfxsize=.48\textwidth
\epsfbox{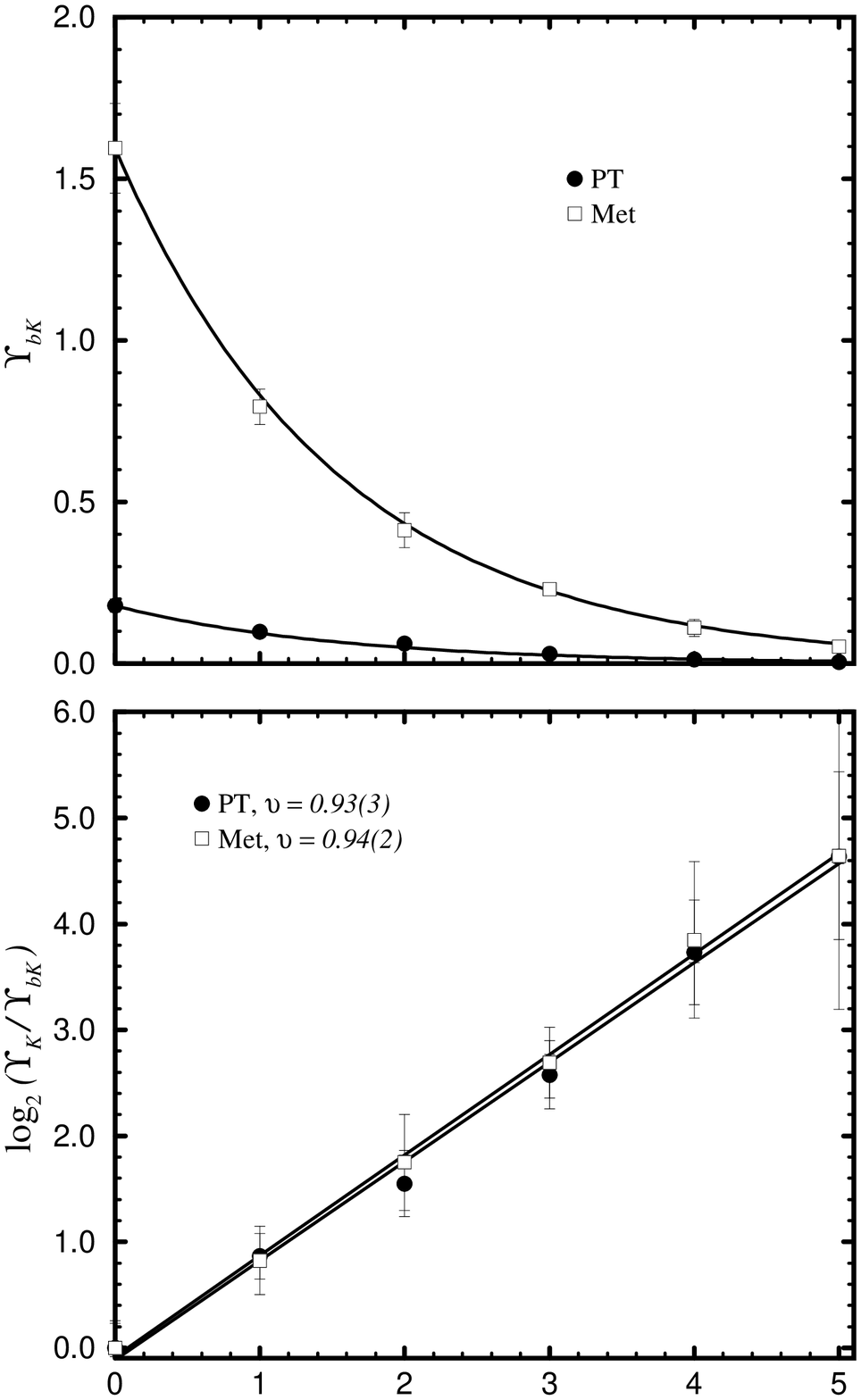}
\end{figure}

{\footnotesize
{\bf Fig. 5:} The upper panel shows the decay behavior of $\Upsilon_{bK}$
(in units of $\varepsilon^2$)
as a function of $\log_2(b)$ for PT and Met, at the temperature of the 
melting peak
of the heat capacity, $k_BT_m/\varepsilon=0.282$. From Eq. (\ref{upa1}), we 
plot $\log_2{(\Upsilon_{K}/\Upsilon_{bK})}$ vs. $\log_2{(b)}$, to extract the 
value of the exponent $\upsilon$ (the slope of the linear fit). We have 
found $\upsilon=0.93\pm0.03$ for PT, and $\upsilon=0.94\pm0.02$ for Met. 
The straight lines are the best linear fits of the data points.}
\vspace{.5cm}

We have stated that the simulation can be considered effectively 
ergodic when $\Upsilon_K$ is indistinguishable from zero. 
In Fig. 4 we have plotted $\Upsilon_{bK}$ and its statistical 
error as a function
of
$\log_2(b)$ for $R_c=2.5$ and $2\sigma$. For $R_c=2\sigma$ the 
crossing point of $\Upsilon_{bK}$ and its error is at $bK\simeq16\times10^4$. 
For $R_c=2.5\sigma$ the crossing point is at a $bK>64\times10^4$. We can 
conclude that for $k_BT_h/\varepsilon=0.393$ and $R_c=2\sigma$ the 
simulation can be considered effectively ergodic after $16\times10^4$ Met 
passes.

Once a constraining radius is chosen, PT and Jw simulations require the
highest temperature $T_h$ be chosen so that Met is ergodic.  For a given
$R_c$, the extent of ergodicity can be tested using the same metric that
has been used for determining the optimum value of $R_c$, but by varying
the temperature.
For the parameters $k_BT_h/\varepsilon=0.393$ and $R_c=2\sigma$ the simulation 
is ergodic even at very short sequence lengths.  We have 
found that for $k_BT_h/\varepsilon<0.393$ the simulations are not ergodic. 
To be sure that the parameters are appropriate, we have performed a short PT 
simulation ($10^4$ passes, ten PT passes consists of nine Met passes
plus an exchange attempt) with 40 equally spaced temperatures in the 
range $k_BT/\varepsilon=$[0.028,0.393] in order to obtain a first estimate of the position of the 
melting and boiling temperature regions. The boiling peak in the specific 
heat appears to be located at a higher temperature than $k_BT/\varepsilon=$
0.393. Moreover, 
the value of $C_V$ at $k_BT/\varepsilon=0.393$ is about one-half the value of $C_V$ at 
the temperature of the melting peak $k_BT_m/\varepsilon=0.282$. From 
these results we
feel it is safe to choose $R_c=2\sigma$ and $k_BT_h/\varepsilon=0.393$ for
the calculations that follow.

We now illustrate the convergence characteristics of $\Upsilon_K$
when we increase the total time scale of the calculation by a
factor $b$.  We illustrate this behavior using a PT simulation of
LJ$_{13}$, and we focus on results at the temperature of the melting
peak in the heat capacity ($k_BT_m/\varepsilon=0.282$).  
We choose this temperature, because from
experience 
\cite{juan1,juan2,juan3}
we know the statistical fluctuations are large at the melting
heat capacity maximum.  The large statistical fluctuations make it
possible to emphasize the behavior of $\Upsilon_K$.
We have run the PT simulation at 40 equally spaced temperatures in the 
range $k_BT/\varepsilon=$ [0.028,0.393]. The initial warmup time has been 
set to $10^4$ 
Met passes, followed by $2\times10^4$ PT passes. Following the warm-up
period, we perform 
simulations of $10^5$, $2\times10^5$, $4\times10^5$, $8\times10^5$, 
$16\times10^5$, and $32\times10^5$ PT passes.  In each case the initial 
configuration has been taken to be the last configuration of the previous run.
The 
output of the simulation are sequences of the potential energy. 
$\Upsilon_K$ has been determined in the same way
as in the calculation of the high temperature parameters (presented in
Fig. 3 and Fig. 4). The data points appearing in the upper panel of Fig. 5 are 
the mean value over twenty independent calculations of the slope of 
$1/Y_K(\omega_n)$. In the lower panel of Fig. 5 we have plotted 
$\log_2{(\Upsilon_{K}/\Upsilon_{bK})}$ as a function of 
$\log_2{(b)}$, where $K=10^4$ 
and $b=1,2,4,\dots,32$. The slope of the linear fit is the exponent 
$\upsilon$, according to Eq. (\ref{upa1}). At the temperature of the 
melting peak, $\upsilon=0.93\pm0.03$. 

It is of interest to perform a similar study of the behavior of
$\Upsilon_K$ as a function of the time scaling for an Met calculation. 
We have taken the final configuration of the PT simulation at
$k_BT_m/\varepsilon=0.282$ as an initial configuration, and we have 
performed a simple Met simulation at that melting temperature.  
A graph of $\Upsilon_{bK}$ and $\log_2{(\Upsilon_{K}/\Upsilon_{bK})}$ 
as a function of $\log_2{(b)}$ for Met is also presented in Fig. 5. 
From the upper panel of Fig. 5, it is evident that Met results are not
ergodic within the same scaled time as the PT results.  It is also
evident that the power law exponent for both Met and PT 
are not distinguishable. 
Similar studies of the power law using the Jw method also give the same
exponent.  Neither an increase in the number of temperatures nor
changing the distribution of temperatures in both Jw and PT simulations
has any effect on the calculated exponent.

\begin{figure}
\epsfxsize=.48\textwidth
\epsfbox{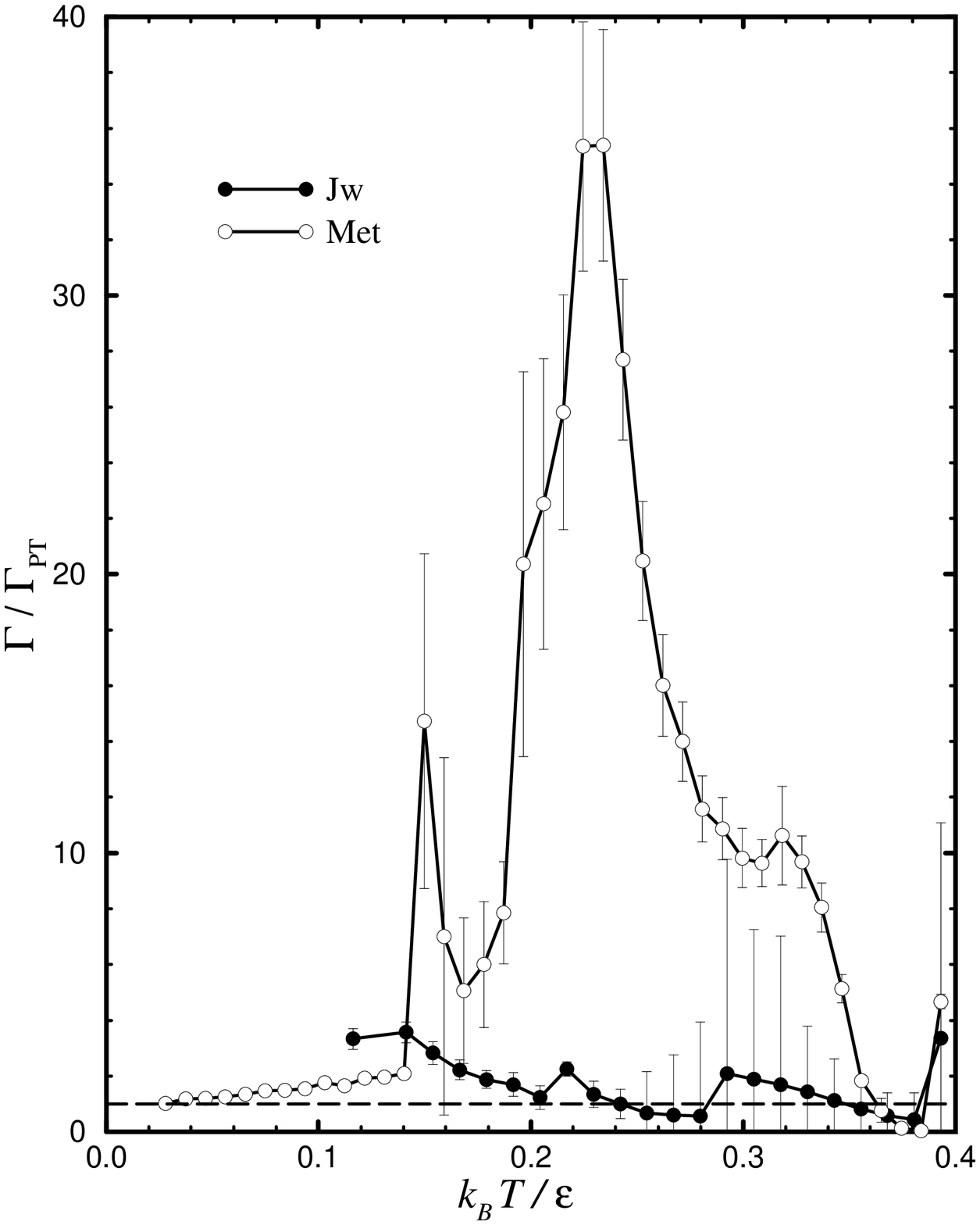}
\end{figure}

{\footnotesize
{\bf Fig. 6:} Comparison of the Met and Jw diffusion coefficients
with the PT diffusion coefficient as a function of the reduced temperature. 
The dashed line represents equivalence between methods.}
\vspace{.5cm}

By using the results to compare the relative
efficiencies of Met, Jw and PT simulations for the LJ$_{13}$ system.
We have found that
PT and Jw simulations can be considered ergodic if the run length is on the 
order of $2\times10^5$ passes, while Met simulations that are initialized 
from configurations
generated from an ergodic PT study are ergodic when the total run length
consists of $2\times10^6$ passes or more.

In order to compare approaches, 
we have calculated $\Gamma$ as a function of the reduced temperature, 
for the three methods. The comparison of diffusion coefficients from different
algorithms has also been used by Andricioaei and
Straub
\cite{straub}.
The comparison of Jw and Met 
with PT is presented in Fig. 6. The Jw and PT simulations are found
to have comparable efficiencies using $\Gamma$ as a measure for all
calculated temperatures.  At intermediate temperatures, Met is
significantly less efficient.  We have chosen to truncate the Jw study
at $k_BT/\varepsilon=0.12$.  For temperatures below $k_BT/\varepsilon=0.12$,
Jw simulations require significant effort, because a large set of
external distributions must be generated.  Because at temperatures below
$k_BT/\varepsilon=0.12$ LJ$_{13}$ is dominated by structures close to the
lowest energy icosahedral isomer, we expect the Jw and PT methods to
have similar efficiencies (as measured by $\Gamma$) for all
temperatures.

\section{Conclusions}

In this paper we have presented a study of the approach to
the ergodic limit in MC simulations. In all the cases examined, 
the behavior of the MC metric $d_k$ can be approximated by Eq. (\ref{d}), 
and the behavior of $\Upsilon_{bK}$ satisfies Eq. (\ref{upa}). Because the 
exponent $\upsilon$ is smaller than one for all the cases studied, the 
dependence of the non-diffusive contributions on $d_k$ is 
weaker (in the sense of Appendix A) than the diffusive contributions. 
The assumption on which we have built the stochastic model 
have been verified numerically for a system having a sufficiently
complex potential surface to be viewed as prototypical of a large set of
many-particle systems.

The MC metric used in this work appears to be a valuable tool to 
study the ergodicity properties of MC simulations. The non-ergodic components
of the MC metric enable the prediction of the minimum length 
a MC simulation must have in order to be considered ergodic. 
The comparison of $\Gamma$ from 
different algorithms gives a reasonable estimate of their relative
efficiencies.

From the study of the melting region of 13 particle clusters, we have 
found that the exponent $\upsilon$ depends both on the method used and the 
nature of the potential energy function. We have performed calculations,
not discussed in this work, where the functional form of the potential
energy is modified.  These studies have shown $\upsilon$ to be dependent
on the details of the potential. We have not found the exponent $\upsilon$ 
to be a strong function of method.  Although PT and Met have
significantly different efficiencies as measured by their relative
diffusion coefficients, $\upsilon$ is nearly the same in the two
methods.  The difference in the decay of $\Upsilon_K$ appears to be
dominated by the coefficient in Eqs.(\ref{upa}) and (\ref{upa1})
rather than the exponent.

As discussed in the text, parallel tempering and J-walking studies of
many-particle systems must have an initial high temperature 
component that is chosen so that a Met simulation is known to be ergodic.
For cluster simulations that require an external constraining potential
to define the cluster, the radius of the constraining potential must be
carefully chosen in order to achieve ergodic results.  We have found the
metric and associated decay laws developed in this work to be a
particularly valuable method of choosing these initial parameters in
both parallel tempering and J-walking simulations.

We also remark that the metric introduced here may be a more
sensitive probe of ergodicity than may be required in some applications. 
For example in previous J-walking studies\cite{clusters} of 
the 13-particle Lennard-Jones cluster,
the heat capacity curve determined with a constraining radius of 4$\sigma$
is nearly indistinguishable from the curve obtained with a constraining
radius of 2$\sigma$.  From the results of this work, we know the initial
high temperature walk is not ergodic when a constraining 
radius of 4$\sigma$ is used.  It is striking that the non-ergodicity as
measured by the energy metric is not apparent in the heat capacity
curve.

We have constructed a metric based on an ensemble of MC trajectories. 
By using an ensemble we attempt to cover sufficient portions of space
so that all components are accessible.  In practice only a finite subset of
a full ensemble can be included, and it is always possible that
components of space are missed.  In such a case $\Upsilon_K$ may decay
to zero numerically within the subspace, and the behavior may give
misleading evidence that the simulation is ergodic.  Because 
components of space may be missed in any finite simulation, it is impossible to
guarantee ergodicity.  It is hoped by using a sufficiently large 
ensemble of trajectories to define the metric, the possibility of
missing components is minimized.

\section*{Acknowledgments}

We would like to thank Dr. O. Osenda for helpful comments.
This work has been supported in part by the National Science Foundation under
grant numbers CHE-9714970 and CDA-9724347.
This research has 
been supported in
part by the Phillips Laboratory, Air Force Material Command, USAF,
through the use of the MHPCC under cooperative agreement number
F29601-93-0001.  The views and conclusions contained in this document
are those of the authors and should not be interpreted as necessarily
representing the official policies or endorsements, either expressed or
implied, of Phillips Laboratory or the U.S. Government.

\appendix
\section{Weak dependence of the non-diffusive contributions}

We have considered two overall time scales for a MC simulation. 
Properties calculated at short times (labeled $k$ in the discrete case)
provide information about each step of the MC process, and properties
averaged over the total simulation time (labeled $K$ in the discrete
case) give information about the approach to ergodic behavior.  When $K$
is sufficiently short we have
both diffusive and non-diffusive contributions as a function of $k$.
In this Appendix we explain the relative time dependence of the diffusive and 
non-diffusive contributions to the autocorrelation function.

It has been assumed that the autocorrelation function 
Eq. (\ref{correl}) can be expressed as the sum of diffusive terms plus 
non-diffusive terms, i.e.

\begin{equation}
\label{h1}
\kappa(t,t')= \kappa_d(t,t') + \sum_{\ell=\lambda+1}^{\Lambda}\kappa_{nd\,,\,\ell}(t,t')\,\,,
\end{equation}

\noindent where

\begin{eqnarray}
\kappa_d(t,t')&=&\frac{\Gamma_0+\Gamma_1+\Gamma_2+\ldots+\Gamma_{\lambda
}}
{t_>}\label{h2}\\
\kappa_{nd\,,\,\ell}(t,t')&=& \frac{\Gamma_{\ell}}{\tau_{\ell}} \, f_{\ell}\left(\frac{t_{\ell}^*}{\tau_{\ell}}\right)\label{h3}\,\,.
\end{eqnarray}

Increasing the time variables by a factor $b>1$, such that 
$\tau_{\lambda}\ll bt_>\ll\tau_{\lambda+1}$, with $\lambda\geq1$, 
we can study the relative variations of each contribution to the correlation 
function, diffusive and non-diffusive (labeled by $\ell>\lambda$).  In this
Appendix we only consider values of $b$ such that
the transformation $t\to bt$ does not increase the time scale beyond the
local correlation time.  In Appendix B values of $b$ are considered that
do cross such time scales.

By relative variations we mean

\begin{eqnarray}
\Delta_d(t,t';b) &=&\left|  \frac{\kappa_d(bt,bt')-\kappa_d(t,t')}
{\kappa_d(t,t')}\right|\label{h4}\\
\Delta_{nd\,,\,\ell}(t,t';b) &=&\left|  \frac{\kappa_{nd\,,\,\ell}(bt,bt')
-\kappa_{nd\,,\,\ell}(t,t')}{\kappa_{nd\,,\,\ell}(t,t')}
\right|\label{h5}\,\,.
\end{eqnarray}

\noindent The relative variation of each non-diffusive contribution is

\begin{equation}
\label{h7}
\Delta_{nd\,,\,\ell}(t,t';b) = \left|1-\frac1{b^2} \frac{\int_0^{bt}\! dt_1
\int_0^{bt'}\! dt_2 \,f_{\ell}\left(\frac{|t_1-t_2|}{\tau_{\ell}}\right)}{\int_0^t\! dt_1
\int_0^{t'}\! dt_2 \,f_{\ell}\left(\frac{|t_1-t_2|}{\tau_{\ell}}\right)}\right|\,.
\end{equation}

\noindent whereas the relative variation of the diffusive contribution is

\begin{equation}
\label{h6}
\Delta_d(t,t';b) = 1-\frac1b\,\,,
\end{equation}

\noindent If $\Delta_d(t,t';b)>\Delta_{nd\,,\,\ell}(t,t';b)$ for 
all pair of times $t$ and $t'$ and for all $b>1$ such that 
$bt_>\ll\tau_{\ell}$, we say that the  non-diffusive contributions are weaker 
than the diffusive contribution in their dependence on $t$. We explore, 
in the remainder of this appendix, the properties $f_{\ell}$ must have in 
order that the inequality 
$\Delta_d(t,t';b)>\Delta_{nd\,,\,\ell}(t,t';b)$
is satisfied.

\vspace{.5cm}
{\bf Lemma:} If the function $H_{\ell}(t;\tau)$

\begin{equation}
\label{lemma1}
H_{\ell}(t;\tau) = \int_0^t\!\!dt'\,\,f_{\ell}
\left(\frac{t'}{\tau}\right)>0\,\,
\end{equation}

\noindent satisfies the inequality

\begin{equation}
\label{lemma2}
H_{\ell}(t;\tau) >t\,f_{\ell} \left(\frac{t}{\tau}\right)\,\,\forall\,\,t\,
{\rm and}\,\tau\,\,,
\end{equation}

\noindent then, $H_{\ell}(t;\tau)$ is an increasing function of $\tau$.

\vspace{.5cm}
{\bf Demonstration:} For $\ell$ and $t$ fixed, the function $H_{\ell}(t,\tau)$ 
evaluated in $\tau'$ is

\begin{eqnarray}
H_{\ell}(t;\tau')&=&\int_0^t\!\!dt'\,\,f_{\ell}\left(\frac{t'}{\tau'}
\right)\\
&=&\int_0^t\!\!dt'\,\,f_{\ell}\left(\frac{\tau t'}{\tau'\tau}\right)\\
&=&\frac{\tau'}{\tau} \int_0^{\tau t/\tau'}\!\!du\,\,f_{\ell}
\left(\frac{u}{\tau}\right)\\
&=&\frac{\tau'}{\tau} H_{\ell}(\tau t/\tau';\tau) \label{n0}\,\,,
\end{eqnarray}

\noindent then, for $\Delta\tau>0$

\end{multicols}
\renewcommand{\thesection}{\Alph{section}}
\renewcommand{\theequation}{\thesection\arabic{equation}}

\begin{eqnarray}
\frac{H_{\ell}(t;\tau+\Delta\tau)-H_{\ell}(t;\tau)}{\Delta\tau}&=& 
\frac1{\Delta\tau} \left\{\frac{\tau+\Delta\tau}{\tau} \int_0^
{\tau t/(\tau+\Delta\tau)}\!\!dt'\,\,f_{\ell}\left(\frac{t'}{\tau}\right) -
\int_0^t\!\!dt'\,\,f_{\ell}\left(\frac{t'}{\tau}\right)\right\}\\
&=&\frac1{\Delta\tau} \left\{
\frac{\Delta\tau}{\tau}\,\int_0^{\tau t/(\tau+\Delta\tau)}
\!\!dt'\,\,f_{\ell}\left(\frac{t'}{\tau}\right) -
\int_{\tau t/(\tau+\Delta\tau)}^t\!\!dt'\,\,f_{\ell}\left(\frac{t'}
{\tau}\right)
\right\}\\
&=&\frac1{\Delta\tau} \left\{
\frac{\Delta\tau}{\tau}\,\int_0^{\tau t/(\tau+\Delta\tau)}\!
\!dt'\,\,f_{\ell}\left(\frac{t'}{\tau}\right) -
\frac{t\Delta\tau}{\tau+\Delta\tau}\,f_{\ell}\left(\frac{t^*}{\tau}\right) 
\right\}\label{app1}\,\,,
\end{eqnarray}

\begin{multicols}{2}

\noindent where $t^*\in[t\tau/(\tau+\Delta\tau),t]$. In the limit 
$\Delta\tau\to0$, and by virtue of the continuity of $f_{\ell}$, 
the derivative takes the form

\begin{equation}
\label{derivada}
\frac{\partial H_{\ell}(t;\tau)}{\partial\tau}=\frac1{\tau}
\left\{H_{\ell}(t;\tau)-tf_{\ell} \left(\frac t{\tau}\right)\right\}
\label{der}\,\,.
\end{equation}

\noindent Then, $\partial H_{\ell}(t;\tau)/\partial\tau>0$, and 
$H_{\ell}(t;\tau)$ is an increasing function of $\tau.\;\;\Box$ 

Here we have presented the two first conditions 
$f_{\ell}$ must have, namely Eqs. (\ref{lemma1}) and (\ref{lemma2}). 
From Eq. (\ref{schwartz}) $f_{\ell}(0)$ is a global maximum, and
the memory functions must have a positive peak at zero. The area below that 
peak must be sufficiently large to satisfy Eq. (\ref{lemma1}). 
Moreover, $f_{\ell}(0)$ must be sufficiently large to 
satisfy Eq. (\ref{lemma2}), even at points where
$f_{\ell}(t/\tau)$ is a local maximum. Then, to satisfy this Lemma, 
we need a memory function with a sufficiently large global maximum at $t=0$.

\vspace{.5cm}
{\bf Corollary:} Suppose $H_{\ell}(t;\tau_{\ell})>tf_{\ell}(t/\tau_{\ell})$. If $b>1$, then $0<\Delta_{nd\,,\,\ell}(t,t';b)<1$ for all pair of times $t$ and $t'$.
\vspace{.5cm}

{\bf Demonstration:} Under the change of scale in time $t\to bt$, 
$\kappa_{nd\,,\,\ell}(t,t')$ can be written

\end{multicols}

\begin{eqnarray}
\kappa_{nd\,,\,\ell}(bt,bt')&=&\frac1{b^2tt'}\int_0^{bt}\!\! dt_1
\int_0^{bt'}\!\! dt_2 \,\,\frac1{\tau_{\ell}}f_{\ell}
\left(\frac{|t_1-t_2|}{\tau_{\ell}}\right)\\
&=&\frac1{tt'}\int_0^t\!\! dt_1
\int_0^{t'}\!\! dt_2 \,\,\frac1{\tau_{\ell}}f_{\ell}
\left(\frac{b|t_1-t_2|}{\tau_{\ell}}\right)\,\,,
\end{eqnarray}

\noindent then, the quotient 
$\kappa_{nd\,,\,\ell}(bt,bt')/\kappa_{nd\,,\,\ell}(t,t')$ is

\begin{eqnarray}
\frac{\kappa_{nd\,,\,\ell}(bt,bt')}{\kappa_{nd\,,\,\ell}(t,t')} &=& 
\frac{\int_0^t\!\! dt_1\left\{
\int_0^{t_1}\!\! dt \,\,f_{\ell}\left(\frac{t}{\tau_{\ell}/b}\right)
+ \int_0^{t_>-t_1}\!\! dt \,\,f_{\ell}\left(\frac{t}{\tau_{\ell}/b}\right)
\right\}} {\int_0^t\!\! dt_1
\left\{
\int_0^{t_1}\!\! dt \,\,f_{\ell}\left(\frac{t}{\tau_{\ell}}\right)+
\int_0^{t_>-t_1}\!\! dt \,\,f_{\ell}\left(\frac{t}{\tau_{\ell}}\right)
\right\}}\\
&=&\frac{\int_0^t\!\! dt_1\left\{H_{\ell}(t_1;\tau_{\ell}/b)
+H_{\ell}(t_>-t_1;\tau_{\ell}/b)\right\}}
{\int_0^t\!\! dt_1\left\{H_{\ell}(t_1;\tau_{\ell})
+H_{\ell}(t_>-t_1;\tau_{\ell})\right\}}\,\,.
\label{n1}
\end{eqnarray}

\begin{multicols}{2}

\noindent By Eq. (\ref{lemma1}), $H_{\ell}(t;\tau)>0\,\,\forall$ $t$ and 
$\tau$. By the {\bf Lemma} the numerator is smaller than the denominator. 
Then $0<\kappa_{nd\,,\,\ell}(bt,bt')/\kappa_{nd\,,\,\ell}(t,t')<1$ and 
then, $0<\Delta_{nd\,,\,\ell}(t,t';b)<1.\;\;\Box$

\vspace{.5cm}
{\bf Theorem:} Suppose that $b>1$ is such that 
$\tau_{\ell-1}\ll bt_>\ll\tau_{\ell}$, 
 $H_{\ell}(t;\tau_{\ell})>tf_{\ell}(t/\tau_{\ell})$, and
all $f_{\ell}$ satisfy the Lipschitz condition 
\cite{fomin}
(for all
closed interval ${\cal A}$ exists a real positive number $C_{\ell}$ such that

\begin{equation}
\left|f_{\ell}(x) - f_{\ell}(y)\right|\leq C_{\ell}\,\left|x-y\right|
\label{lipschitz}
\end{equation}

for all $x$ and $y$ in ${\cal A}$). 
Then $\Delta_{nd\,,\,\ell}(t,t';b)<\Delta_d(t,t';b)$ 
if and only if $f_{\ell}$ is non-negative in the interval $[0,t_>)$.

\vspace{.5cm}
{\bf Demonstration:} If $\Delta_{nd\,,\,\ell}(t,t';b)<\Delta_d(t,t';b)$,
then

\begin{eqnarray}
1-\frac1b &>& 1-\frac1{b^2} \frac{\int_0^{bt}\! dt_1
\int_0^{bt'}\! dt_2 \,\,f_{\ell}\left(\frac{|t_1-t_2|}{\tau_{\ell}}\right)}
{\int_0^t\! dt_1
\int_0^{t'}\! dt_2 \,\,f_{\ell}\left(\frac{|t_1-t_2|}{\tau_{\ell}}\right)}\\
1&<&\frac1b \frac{\int_0^{bt}\! dt_1
\int_0^{bt'}\! dt_2 \,\,f_{\ell}\left(\frac{|t_1-t_2|}{\tau_{\ell}}\right)}
{\int_0^t\! dt_1
\int_0^{t'}\! dt_2 \,\,f_{\ell}\left(\frac{|t_1-t_2|}{\tau_{\ell}}\right)}
\label{ineq}
\end{eqnarray}

\noindent where the operations to reach Eq. (\ref{ineq}) are valid by 
using {\bf Corollary}. Then

\end{multicols}

\begin{eqnarray}
0&<& \int_0^{bt}\!\! dt_1
\int_0^{bt'}\!\! dt_2\,\, \frac1b\,f_{\ell}\left(\frac{|t_1-t_2|}{\tau_{\ell}}
\right)-
\int_0^{t}\!\! dt_1
\int_0^{t'}\!\! dt_2 \,\,f_{\ell}\left(\frac{|t_1-t_2|}{\tau_{\ell}}\right)\\
0&<& \int_0^{t}\!\! dt_1
\int_0^{t'}\!\! dt_2 \left\{b\,f_{\ell}\left(\frac{b\,|t_1-t_2|}{\tau_{\ell}}
\right)-
\,f_{\ell}\left(\frac{|t_1-t_2|}{\tau_{\ell}}\right)\right\}\\
0&<& \int_0^{t_<}\!\! dt_1\left\{  
\int_0^{t_1}\!\! dt_2 \left[b\,f_{\ell}\left(\frac{b\,(t_1-t_2)}{\tau_{\ell}}
\right)-
\,f_{\ell}\left(\frac{t_1-t_2}{\tau_{\ell}}\right)\right]+
\int^{t_>}_{t_1}\!\! dt_2 \left[b\,f_{\ell}\left(\frac{b\,(t_2-t_1)}
{\tau_{\ell}}\right)-
\,f_{\ell}\left(\frac{t_2-t_1}{\tau_{\ell}}\right)\right]
\right\}\\
0&<& \int_0^{t_<}\!\! dt_1\left\{  
\int_0^{t_1}\!\! dt \left[b\,f_{\ell}\left(\frac{bt}{\tau_{\ell}}\right)-
\,f_{\ell}\left(\frac{t}{\tau_{\ell}}\right)\right]
+
\int_0^{t_>-t_1}\!\! dt \left[b\,f_{\ell}\left(\frac{bt}{\tau_{\ell}}\right)-
\,f_{\ell}\left(\frac{t}{\tau_{\ell}}\right)\right]
\right\}\\
0&<& \int_0^{t_<}\!\! dt_1\left\{  
\int_0^{bt_1}\!\! dt \,\,f_{\ell}\left(\frac{t}{\tau_{\ell}}\right)-
\int_0^{t_1}\!\! dt \,\,f_{\ell}\left(\frac{t}{\tau_{\ell}}\right)
+
\int_0^{b(t_>-t_1)}\!\! dt \,\,f_{\ell}\left(\frac{t}{\tau_{\ell}}\right)-
\int_0^{t_>-t_1}\!\! dt \,\,f_{\ell}\left(\frac{t}{\tau_{\ell}}\right)
\right\}\\
0&<& \int_0^{t_<}\!\! dt_1\left\{  
\int_{t_1}^{bt_1}\!\! dt \,\,f_{\ell}\left(\frac{t}{\tau_{\ell}}\right)
+
\int_{t_>-t_1}^{b(t_>-t_1)}\!\! dt \,\,f_{\ell}\left(\frac{t}{\tau_{\ell}}
\right)\right\}\,\,.\label{k1}
\end{eqnarray}

\noindent Using the intermediate 
value theorem, \cite{spivak} we have

\begin{eqnarray}
\int_t^{bt}\!\!dt'\,\,f_{\ell}\left(\frac{t'}{\tau_{\ell}}
\right)&=&(b-1)\,t \,f_{\ell}\left(\frac{t^*(t)}{\tau_{\ell}}\right)\\
&=&(b-1)\,t\,f_{\ell}\left(\frac{t}{\tau_{\ell}}\right) + (b-1)\,t\,
\left[f_{\ell}\left(\frac{t^*(t)}{\tau_{\ell}}\right) - 
f_{\ell}\left(\frac{t}{\tau_{\ell}}\right)\right]\label{propx}\,\,,
\end{eqnarray}

\noindent where $t^*(t)\in[t,bt]$. Let be $t^*_{\alpha}(t)$ and
$t^*_{\beta}(t)$ the values at 
which the intermediate value theorem is satisfied, in the intervals $[t,bt]$ 
and $[t_>-t,b(t_>-t)]$ respectively

\begin{eqnarray}
(b-1)\,t\,f_{\ell}\left(\frac{t^*_{\alpha}(t)}{\tau_{\ell}}\right) &=&
\int_t^{bt}\!\!dt'\,\,f_{\ell}\left(\frac{t'}{\tau_{\ell}}\right)\\
(b-1)\,(t_>-t)\,f_{\ell}\left(\frac{t^*_{\beta}(t)}{\tau_{\ell}}\right) 
&=&
\int_{t_>-t}^{b(t_>-t)}\!\!dt'\,\,f_{\ell}\left(\frac{t'}
{\tau_{\ell}}\right)\,\,,
\end{eqnarray}

\noindent then, the remainder can be written as

\begin{equation}
\label{rem}
R_{\ell}(t_<,t_>;b) = \int_0^{t_<}\!\!dt\,\,\left\{t\,\left[f_{\ell}
\left(\frac{t^*_{\alpha}(t)}{\tau_{\ell}}\right) 
- f_{\ell}\left(\frac{t}{\tau_{\ell}}\right)\right]+(t_>-t)\,\left[f_{\ell}
\left(\frac{t^*_{\beta}(t)}{\tau_{\ell}}\right) 
- f_{\ell}\left(\frac{t_>-t}{\tau_{\ell}}\right)\right]\right\}\,\,.
\end{equation}

\noindent By the Lipschitz condition, we have that

\begin{eqnarray}
R_{\ell}(t_<,t_>;b) &\leq& \int_0^{t_<}\!\!dt\,\,\left\{t\,\left|f_{\ell}
\left(\frac{t^*_{\alpha}(t)}{\tau_{\ell}}\right) 
- f_{\ell}\left(\frac{t}{\tau_{\ell}}\right)\right|+(t_>-t)\,\left|f_{\ell}
\left(\frac{t^*_{\beta}(t)}{\tau_{\ell}}\right) 
- f_{\ell}\left(\frac{t_>-t}{\tau_{\ell}}\right)\right|\right\}\\
&<& \int_0^{t_<}\!\!dt\,\,\left\{t\,C_{\ell}\,\left|
\frac{t^*_{\alpha}(t)-t}{\tau_{\ell}}\right|+(t_>-t)\,C_{\ell}\,\left|
\frac{t^*_{\beta}(t)-(t_>-t)}{\tau_{\ell}}\right|\right\}\\
&<& \frac{C_{\ell}}{\tau_{\ell}}\,\int_0^{t_<}\!\!dt\,\,\left\{t\,\left|
bt-t\right|+(t_>-t)\,\left|
b(t_>-t)-(t_>-t)\right|\right\}\\
&<& \frac{C_{\ell}}{\tau_{\ell}}\,(b-1)\,\int_0^{t_<}\!\!dt\,\,\left[t^2
+(t_>-t)^2\right]\\
&<& \frac{C_{\ell}}{\tau_{\ell}}\,(b-1)\,\left(\frac23\,t_<^3+t_<t_>\,
(t_>-t_<)\right)\\
&<& \frac23\,t_>^3\,\frac{C_{\ell}}{\tau_{\ell}}\,(b-1)\label{cota}\,\,,
\end{eqnarray}

\noindent where $C_{\ell}$ is a suitable positive real constant. Using Eqs. 
(\ref{propx}) and (\ref{rem}) in Eq. (\ref{k1}) we have

\begin{eqnarray}
0&<& \int_0^{t_<}\!\! dt\,\,(b-1)\,\left\{ t\,f_{\ell}\left(\frac{t}
{\tau_{\ell}}\right) + (t_>-t)\,f_{\ell}\left(\frac{t_>-t}{\tau_{\ell}}
\right)\right\}+(b-1)\,R_{\ell}(t_<,t_>;b) \\
0&<& \int_0^{t_<}\!\! dt\,\,t\,f_{\ell}\left(\frac{t}{\tau_{\ell}}\right) +
\int_{t_>-t_<}^{t_>}\!\! dt\,\,t\,f_{\ell}\left(\frac{t}{\tau_{\ell}}\right)+
R_{\ell}(t_<,t_>;b)\\
0&<& \int_0^{t_<}\!\! dt\,\,t\,f_{\ell}\left(\frac{t}{\tau_{\ell}}\right) +
\int_0^{t_>}\!\! dt\,\,t\,f_{\ell}\left(\frac{t}{\tau_{\ell}}\right)-
\int_0^{t_>-t_<}\!\! dt\,\,t\,f_{\ell}\left(\frac{t}{\tau_{\ell}}\right)
+R_{\ell}(t_<,t_>;b)\\
0&<& F_{\ell}(t_<) + F_{\ell}(t_>) - F_{\ell}(t_> - t_<)+
\frac23\,t_>^3\,\frac{C_{\ell}}{\tau_{\ell}}\,(b-1)
\label{xx}\,\,,
\end{eqnarray}

\noindent where

\begin{equation}
F_{\ell}(t) = \int_0^t\!\! dt'\,\,t'\,f_{\ell}\left(\frac{t'}{\tau_{\ell}}
\right)\,\,,
\end{equation}

\noindent is a continuous and differentiable function of $t$. 
The inequality (\ref{xx}) holds for any $b>1$. Suppose that 
$F_{\ell}(t_<) + F_{\ell}(t_>) - F_{\ell}(t_> - t_<)<0$. Then, if
$b$ is such that

\begin{equation}
b = 1 + \frac3{2L}\,\frac{\tau_{\ell}}{t_>^3\,C_{\ell}}\,
\left|F_{\ell}(t_<) + F_{\ell}(t_>) - F_{\ell}(t_> - t_<)\right|\,\,,
\end{equation}

\noindent where $L>2$, we have that

\begin{eqnarray}
0&<& F_{\ell}(t_<) + F_{\ell}(t_>) - F_{\ell}(t_> - t_<)+
\frac1L\,\left|F_{\ell}(t_<) + F_{\ell}(t_>) - F_{\ell}(t_> - t_<)\right|\\
0&<&\frac{L-1}L\,\left[F_{\ell}(t_<) + F_{\ell}(t_>) - F_{\ell}(t_> - t_<)\right]
\end{eqnarray}

\noindent in contradiction with the hypothesis that 
$F_{\ell}(t_<) + F_{\ell}(t_>) - F_{\ell}(t_> - t_<)$ is negative. Then

\begin{equation}
0\leq F_{\ell}(t_<) + F_{\ell}(t_>) - F_{\ell}(t_> - t_<)\label{xxx}\,\,.
\end{equation}

Let us define the function 

\begin{equation}
\Delta F_{\ell}(t) = F_{\ell}(t) - F_{\ell}(t_> - t)\,\,,
\end{equation}

\noindent where $t\in(0,t_>)$. The right derivative at $t=0$ of 
$\Delta F_{\ell}(t)$ is

\begin{eqnarray}
\lim_{\Delta t\to0^+}\frac{\Delta F_{\ell}(\Delta t) - \Delta F_{\ell}(0)}
{\Delta t}&=&\lim_{\Delta t\to0^+}\frac{F_{\ell}(\Delta t)-F_{\ell}(0)
+F_{\ell}(t_>)-F_{\ell}(t_>-\Delta t)}{\Delta t}\\
&=&
\lim_{\Delta t\to0^+}\frac1{\Delta t}
\left\{\int_0^{\Delta t}\!\!dt\,\,t\,f_{\ell}\left(\frac{t}{\tau_{\ell}}
\right)+\int^{t_>}_{t_>-\Delta t}\!\!dt\,\,t\,f_{\ell}\left(\frac{t}
{\tau_{\ell}}\right)
\right\}\\
&=&\lim_{\Delta t\to0^+}\frac1{\Delta t}
\left\{\Delta t \,\,t_1^* f_{\ell}\left(\frac{t_1^*}{\tau_{\ell}}\right) 
+ \Delta t \,\,t_2^* f_{\ell}\left(\frac{t_2^*}{\tau_{\ell}}\right)\right\}
\label{n3}
\end{eqnarray}

\begin{multicols}{2}

\noindent where $t_1^*\in[0,\Delta t]$ and $t_2^*\in[t_>-\Delta t,t_>]$. Thus

\begin{equation}
\label{n5}
\left.\frac{\partial\Delta F_{\ell}(t)}{\partial t}\right|_{t\to 0^+}
= t_>\,f_{\ell}\left(\frac{t_>}{\tau_{\ell}}\right)\,\,.
\end{equation}

\noindent If the right derivative at 0 of $\Delta F_{\ell}(t)$ is 
negative, $\Delta F_{\ell}(t)$ approaches $-F_{\ell}(t_>)$ from below, 
when $t\to0$. There exists a time $0<\tilde{t}<t_>$, such that 
$0> F_{\ell}(t_>) + \Delta F_{\ell}(\tilde{t})$, in contradiction with 
Eq. (\ref{xxx}). Then, $f_{\ell}$ must be non-negative for $t\in(0,t_>)$. 
By the property Eq. (\ref{schwartz}) $f_{\ell}(0)$ must be positive.
This proves that $\Delta_{nd\,,\,1}(t,t';b) < \Delta_d(t,t';b) 
\Rightarrow f_{\ell}(t)\geq0$ for $0 \le t < t_{>}$.
To demonstrate that if $f_{\ell}$ is positive yields 
$\Delta_{nd\,,\,1}(t,t';b)<\Delta_d(t,t';b)$ (i.e. the converse), 
follow the argument 
backwards, from Eq. (\ref{k1}). $\Box$

In conclusion, if the memory functions are positive, 
satisfy the Lipschitz condition, and satisfy the 
condition Eqs. (\ref{lemma1}) and (\ref{lemma2}), 
the non-diffusive 
contributions are more weakly dependent on time than $1/t$. 

The results of the present appendix are valid in the 
limit of a complete 
ensemble. In our numerical experiments only partial samples of the ensemble can
be considered. The memory functions that appear in our numerical calculations 
come from partial mean values of the 
product of discontinuous
functions (every noise process is a discontinuous function). These
memory functions are discontinuous. The behavior of the 
non-diffusive contributions observed in our
numerical experiments is in agreement with these analytic 
(infinite ensemble limit) results. We can infer that there might be
a version of the theorem applied to discontinuous memory functions, but
we have been unable to develop such a theorem.

\section{Consequences of the time scale change in the non-diffusive 
contributions}

In this appendix we show the behavior of the function $f_1$ when its 
correlation time is changed according to $\tau_1\to\tau_{b1}=\tau_1/b$, 
with $b\gg1$; i.e. when the total simulation time is scaled to exceed
the correlation time of the first colored noise process.

We multiply the time variables by a number $b$, such that 
$\tau_1\ll bt_> \ll \tau_2$. We have that the $g_1$ 
process contributes to the autocorrelation function with

\end{multicols}

\begin{eqnarray}
\frac1{b^2tt'}\langle G_1(bt/\tau_{\ell})\,G_1(bt'/\tau_{\ell})
\rangle &=&\frac1{b^2tt'}\, \int_0^{bt_<}\!\! dt_1\int_0^{bt_>}\!\! 
dt_2\,\,\frac1{\tau_1}f_1\left(\frac{|t_1-t_2|}{\tau_1}\right)\\
&=& \frac1{btt'}\,\int_0^{t_<}\!\! dt_1'\int_0^{t_>}\!\! dt_2'\,\,
\frac1{\tau_{b1}}f_1\left(\frac{|t_1'-t_2'|}{\tau_{b1}}\right) \label{eq:6}
\end{eqnarray}

\noindent where $t'=t/b$ and $\tau_{b1}=\tau_1/b$. We want to 
compute this contribution both within the neighborhood $t_1=t_2$ as well
as outside
such a region. To do so, we can split the integral in Eq. (\ref{eq:6}) 
in three parts

\begin{equation}
\frac1{b^2tt'}\langle G_1(bt/\tau_{\ell})\,G_1(bt'/\tau_{\ell})\rangle = 
I_1+I_2+I_3
\end{equation}

\noindent where

\begin{eqnarray}
I_1 &=& \frac1{btt'}\,\int_0^{t_<}\!\! dt_1
\int_0^{\max(0,t_1-\epsilon/2)}\!\! dt_2\,\,\frac1{\tau_{b1}}f_1
\left(\frac{t_1-t_2}{\tau_{b1}}\right)\label{i1}\\
I_2 &=& \frac1{btt'}\,\int_0^{t_<}\!\! dt_1
\int_{\max(0,t_1-\epsilon/2)}^{\min(t_>,t_1+\epsilon/2)}\!\! dt_2\,\,
\frac1{\tau_{b1}}f_1\left(\frac{|t_1-t_2|}{\tau_{b1}}\right)\label{i2}\\
I_3 &=& \frac1{btt'}\,\int_0^{t_<}\!\! dt_1
\int_{\min(t_>,t_1+\epsilon/2)}^{t_>}\!\! dt_2\,\,
\frac1{\tau_{b1}}f_1\left(\frac{t_2-t_1}{\tau_{b1}}\right)\label{i3}
\end{eqnarray}

\noindent with $t_<>\epsilon>0$ 
(observe that the only integral involving $t_1=t_2$ is $I_2$). 
Consider $I_1$. If $t_1<\epsilon/2$ the inner integral is zero. 
Therefore, $t_1$ must be bigger than $\epsilon/2$ and

\begin{equation}
I_1 = \frac1{bt_<t_>}\,\int_{\epsilon/2}^{t_<}\!\! dt_1
\int_0^{t_1-\epsilon/2}\!\! dt_2\,\,\frac1{\tau_{b1}}f_1
\left(\frac{t_1-t_2}{\tau_{b1}}\right)\,\,,
\end{equation}

\noindent which, by virtue of the continuity of $f_1$, can be bounded as
follows

\begin{eqnarray}
\frac1{bt_<t_>}\,\int_{\epsilon/2}^{t_<}\!\! dt_1\,\,
\frac b{\tau_1}\,\left(t_1-\frac{\epsilon}2\right)f_1
\left(\frac{bt_{min}}{\tau_1}\right)
\leq &I_1&\leq \frac1{bt_<t_>}\,\int_{\epsilon/2}^{t_<}\!\! dt_1\,\,
\frac b{\tau_1}\,\left(t_1-\frac{\epsilon}2\right)f_1\left(\frac{bt_{max}}
{\tau_1}\right)\nonumber\\
\frac12\,\frac{(t_<-\epsilon/2)^2}{t_<t_>}\,\frac1{\tau_1}\,f_1
\left( \frac{bt_{min}}{\tau_1}\right)\leq&I_1&\leq
\frac12\,\frac{(t_<-\epsilon/2)^2}{t_<t_>}\,\frac1{\tau_1}\,f_1
\left( \frac{bt_{max}}{\tau_1}\right)
\label{i111}
\end{eqnarray}

\noindent where $t_{max}$ ($t_{min}$) is the time in the interval 
$[\epsilon/2,t_<]$ at which the function $f_1$ 
reaches its maximum (minimum) value. Because $f_1$ is continuous, 
there exists $t_1^*\in[t_{min},t_{max}]$ at which

\begin{equation}
\label{i11}
I_1=
\frac12\,\frac{(t_<-\epsilon/2)^2}{t_<t_>}\,\frac1{\tau_1}\,f_1
\left( \frac{bt_1^*}{\tau_1}\right)\,\,.
\end{equation}

Consider now $I_3$. If $t_1+\epsilon/2>t_>$, the inner integral is zero. 
Therefore, $0<t_1<\min(t_<,t_>-\epsilon/2)$ and

\begin{eqnarray}
I_3&=& \frac1{btt'}\,\int_0^{\min(t_<,t_>-\epsilon/2)}\!\! dt_1
\int_{t_1+\epsilon/2}^{t_>}\!\! dt_2\,\,\frac1{\tau_{b1}}f_1
\left(\frac{t_2-t_1}{\tau_{b1}}\right)\nonumber\\
&=& \frac{\min(t_<,t_>-\epsilon/2)}{t_<t_>}\left[t_>-\frac{\epsilon}2
-\frac12 \min(t_<,t_>-\epsilon/2) \right]\,\frac1{\tau_1}\,f_1\left( 
\frac{bt_3^*}{\tau_1}\right)\,\,,
\label{i33}
\end{eqnarray}

\noindent where $t^*_3\in[t_{min},t_{max}]$, and now $t_{max}$ ($t_{min}$) 
is the time in $[\epsilon/2,t_>]$ at which the function $f_1$ reaches its 
maximum (minimum) value.

\begin{figure}
\epsfxsize=.48\textwidth
\epsfbox{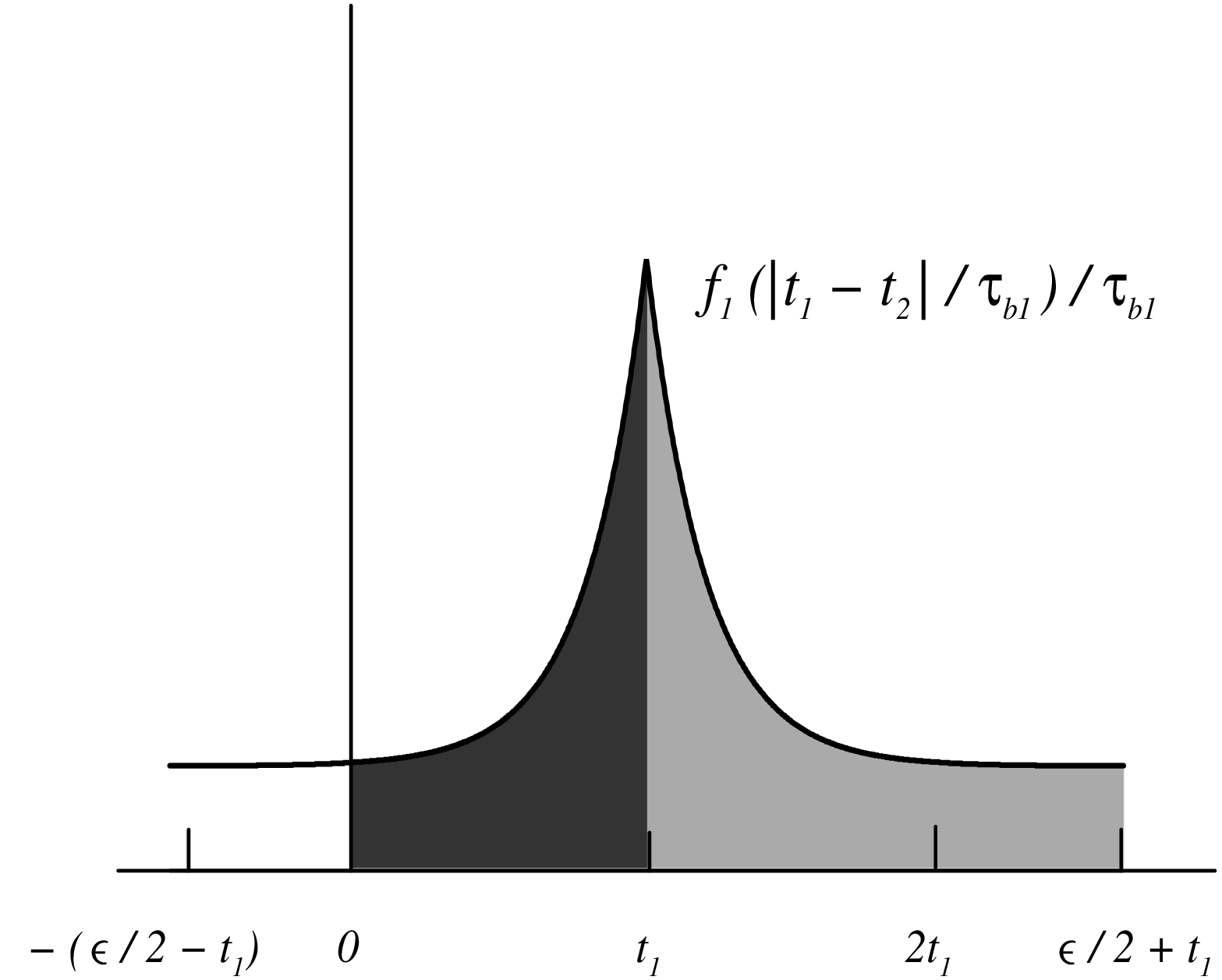}
\end{figure}

{\footnotesize
{\bf Fig. 7:} The area under the curve represents the 
first integral in Eq. (\ref{f1}). The darker piece is 
half of the integral in the interval $[-t_1,t_1]$, the lighter is 
half of the integral in $[-\epsilon/2,\epsilon/2]$.}
\vspace{.5cm}

Let us consider now $I_2$. First observe that for the integral in $t_1$, 
if $0\leq t_1\leq\epsilon/2$, $\max(0,t_1-\epsilon/2)=0$ 
and $\min(t_>,t_1+\epsilon/2)=t_1+\epsilon/2$. 
If $\epsilon/2\leq t_1\leq t_<$ then $\max(0,t_1-\epsilon/2)=
t_1-\epsilon/2$. Then

\begin{equation}
I_2 = \frac1{bt_<t_>}\,\left\{ \int_0^{\epsilon/2}\!\!dt_1 
\int_{0}^{t_1+\epsilon/2}\!\!dt_2\,\,\frac1{\tau_{b1}} f_1
\left(\frac{|t_1-t_2|}{\tau_{b1}}\right) +
\int_{\epsilon/2}^{t_<}\!\!dt_1 
\int_{t_1-\epsilon/2}^{\min(t_>,t_1+\epsilon/2)}\!\!dt_2\,\,
\frac1{\tau_{b1}} f_1\left(\frac{|t_1-t_2|}{\tau_{b1}}\right)\right\}
\label{f1}\,\,.
\end{equation}

\noindent The integral in $t_2$ between 0 and $t_1+\epsilon/2$ can be 
evaluated with the help of Fig. 7

\begin{equation}
\int_{0}^{t_1+\epsilon/2}\!\!dt_2\,\,\frac1{\tau_{b1}} f_1
\left(\frac{|t_1-t_2|}{\tau_{b1}}\right)=\frac12 \,\,
\int_{-\epsilon/2}^{\epsilon/2}\!\!dt\,\,\frac1{\tau_{b1}} 
f_1\left(\frac{|t|}{\tau_{b1}}\right) + \frac12 \,\,
\int_{-t_1}^{t_1}\!\!dt\,\,\frac1{\tau_{b1}} f_1\left(\frac{|t|}
{\tau_{b1}}\right)\,\,.
\label{int1}
\end{equation}

The second integral in $t_1$ can be separated in two parts; the first for 
$\epsilon /2 \leq t_1 \leq \min(t_<,t_>-\epsilon/2)$ and the 
second for $\min(t_<,t_>-\epsilon/2)\leq t_1 \leq t_<$. If 
$t_>-t_<<\epsilon/2$ the second term is zero. Then

\begin{eqnarray}
\int_{\epsilon/2}^{t_<}\!\!dt_1 
\int_{t_1-\epsilon/2}^{\min(t_>,t_1+\epsilon/2)}&&\!\!dt_2\,\,
\frac1{\tau_{b1}} f_1\left(\frac{|t_1-t_2|}{\tau_{b1}}\right) = 
\int_{\epsilon/2}^{\min(t_<,t_>-\epsilon/2)}\!\!dt_1 
\int_{t_1-\epsilon/2}^{\min(t_>,t_1+\epsilon/2)}\!\!dt_2\,\,
\frac1{\tau_{b1}} f_1\left(\frac{|t_1-t_2|}{\tau_{b1}}\right) + 
\nonumber\\ 
&&\;\;\;\;\;\;\;\;\;\;\;\;\;
\Theta\left(\frac{\epsilon}2+t_<-t_>\right)\,\,
\int_{t_>-\epsilon/2} ^{t_<}\!\!dt_1
\int_{t_1-\epsilon/2}^{\min(t_>,t_1+\epsilon/2)}\!\!dt_2\,\,
\frac1{\tau_{b1}} f_1\left(\frac{|t_1-t_2|}{\tau_{b1}}\right)\,\,,
\label{int2}
\end{eqnarray}

\noindent where $\Theta$ is the step function. If 
$t_1\leq\min(t_<,t_> -\epsilon/2)$ then $\min(t_>,t_1+\epsilon/2)=
t_1+\epsilon/2$. The last integral in 
$t_2$ can be rearranged in the same way as 
Eq. (\ref{int1}). Then

\begin{eqnarray}
\int_{\epsilon/2}^{t_<}\!\!dt_1&& 
\int_{t_1-\epsilon/2}^{\min(t_>,t_1+\epsilon/2)}\!\!dt_2\,\,
\frac1{\tau_{b1}} f_1\left(\frac{|t_1-t_2|}{\tau_{b1}}\right) = 
\int_{\epsilon/2}^{\min(t_<,t_>-\epsilon/2)}\!\!dt_1 
\int_{-\epsilon/2}^{\epsilon/2}\!\!dt\,\,\frac1{\tau_{b1}} 
f_1\left(\frac{|t|}{\tau_{b1}}\right) + \nonumber\\ 
&&\;\;\;\;\;\;\;\;\;\;\;\;\;\;\frac12\,
\Theta\left(\frac{\epsilon}2+t_<-t_>\right)\,\,
\int_{t_>-\epsilon/2} ^{t_<}\!\!dt_1\left[
\int_{-\epsilon/2}^{\epsilon/2}\!\!dt\,\,\frac1{\tau_{b1}} 
f_1\left(\frac{|t|}{\tau_{b1}}\right)+
\int_{t_1-t_>}^{t_>-t_1}\!\!dt\,\,\frac1{\tau_{b1}} f_1\left(\frac{|t|}
{\tau_{b1}}\right)\right]\,\,,
\label{int3}
\end{eqnarray}

We can observe that the correlation time $\tau_{b1}$ goes to zero when 
$b$ is increased. The function $(1/\tau_1)\,f_1(bt/\tau_{1})$ 
becomes negligible outside a neighborhood of $t=0$ [observe Eqs. (\ref{i11}) 
and (\ref{i33})]. Equation (\ref{norm}) holds, then, if $b$ is sufficiently 
large, $(1/\tau_{b1})\,f_1(t/\tau_{b1})$ can be considered a delta function. 
The integrals $I_1$ and $I_3$ become zero, and the integrals involving $t=0$ 
in the expression of $I_2$ converge to one. $I_2$ becomes

\begin{equation}
I_2 = \frac1{bt_<t_>} \,\,\left\{ \min\left(t_<,t_>-\frac{\epsilon}2\right) + \Theta\left
(\frac{\epsilon}2+t_<-t_>\right)\, \left(\frac{\epsilon}2+t_<-t_> \right) \right\}=\frac1{bt_>}\,\,,
\label{i22}
\end{equation}

\begin{multicols}{2}

\noindent which is a diffusive contribution to the autocorrelation function. 
The autocorrelation function becomes then

\begin{equation}
\kappa(bt,bt') =
\frac{\Gamma_0+\Gamma_1}{bt_>}+\sum_{\ell=2}^{\Lambda}
\frac{\Gamma_{\ell}}{\tau_{\ell}} \, f_{\ell}
\left(\frac{t_{b\ell}^*(t_>)}{\tau_{b\ell}}\right)\,\,.
\end{equation}

The same argument can be used when $b$ is such that $\tau_2\ll bt_>\ll\tau_3$. After such changes in the time scale, the diffusion coefficient $\Gamma=\Gamma_0+\Gamma_1$ is enlarged, and the non-diffusive contributions are reduced. There is an ultimate scale change, such that $\tau_{\Lambda}\ll bt_>$.
Beyond this maximum time scale the process can be considered diffusive.

\end{multicols}


\begin{references}
\bibitem{citfreq}Assuming the tolerable error to be on the order of 1\%,
we set $0.01 \simeq |[\cot(\omega_n/2)-2/\omega_n]/(2/\omega_n)| =
\omega_n^2/12+{\cal O}(\omega_n^4)$.  Then $n_{max}=bK\sqrt{12}/20\pi$.
\end{references}
\end{document}